\documentclass{article}

\usepackage[english]{babel}
\usepackage[letterpaper,top=2cm,bottom=2cm,left=3cm,right=3cm,marginparwidth=1.75cm]{geometry}

\usepackage{amsmath}
\usepackage{graphicx}
\usepackage[colorlinks=true, allcolors=blue]{hyperref}
\usepackage{url}
\usepackage{amssymb}
\usepackage{caption}
\usepackage{subcaption}
\usepackage{multirow}
\usepackage{siunitx}
\usepackage{tikz-cd}

\usepackage{xcolor}
\newtheorem{theorem}{Theorem}[section]

\newtheorem{proposition}{Proposition}[theorem]

\newcommand{\bigo}{\mathcal{O}}

\newcommand{\bfn}{\mathbf{n}}

\newcommand{\bfq}{\mathbf{q}}

\newcommand{\bfu}{\mathbf{u}}

\newcommand{\bfx}{\mathbf{x}}

\newcommand{\bfA}{\mathbf{A}}
\newcommand{\bfB}{\mathbf{B}}

\newcommand{\bfF}{\mathbf{F}}

\newcommand{\bfQ}{\mathbf{Q}}
\newcommand{\bfR}{\mathbf{R}}
\newcommand{\bfS}{\mathbf{S}}

\newcommand{\barb}{\bar{b}}
\newcommand{\barc}{\bar{c}}

\newcommand{\barh}{\bar{h}}

\newcommand{\barp}{\bar{p}}

\newcommand{\baru}{\bar{u}}

\newcommand{\barx}{\bar{x}}

\newcommand{\real}{\mathbb{R}}

\title{Adaptive, efficient, and scalable  water wave modeling with dispersive hyperbolic systems}
\author{Carlos Muñoz-Moncayo \thanks{Computer, Electrical, and Mathematical Sciences \& Engineering Division, King Abdullah University of Science and Technology, Thuwal 23955, Saudi Arabia, \texttt{carlos.munozmoncayo@kaust.edu.sa}}
\and David I. Ketcheson \thanks{Computer, Electrical, and Mathematical Sciences \& Engineering Division, King Abdullah University of Science and Technology, Thuwal 23955, Saudi Arabia, \texttt{david.ketcheson@kaust.edu.sa}}}

\begin{document}
\date{}
\maketitle

\definecolor{tealblue}{rgb}{0.0, 0.33, 0.71}
\begin{abstract}
    Accurate modeling of tsunamis (such as those generated by landslides) requires capturing
    both wave dispersion in the deep ocean and wave breaking near the shore.
    The shallow water equations are often preferred for working with 
    tsunamis, but neglect dispersion and may be inaccurate in scenarios where dispersive effects are significant.
    In this work we develop an approach that seeks to incorporate the best
    aspects of both hyperbolic and dispersive models, by combining either of two hyperbolic reformulations
    of the Serre-Green-Naghdi equations away from the shore with the non-dispersive shallow water equations near the shore.
    The model is discretized and implemented within the GeoClaw software, and incorporates adaptive mesh refinement
    as well as shared-memory parallelism.  We validate it
    through comparison with benchmarks and real tsunami data.  The results and performance compare
    favorably with the existing dispersive water wave solvers, including a speedup of about 2x relative
    to GeoClaw's existing dispersive solver for a large-scale tsunami simulation.
\end{abstract}

\section{Motivation and goals}
\label{sec:motivation}

The shallow water equations (SWE) are the workhorse of large-scale tsunami
modeling. Derived from the incompressible Euler equations with free
surface under the assumption that the characteristic wavelength is much larger
than the water depth, they capture the essential physics of long-wave
propagation. Crucially, they admit shock wave solutions whose energy dissipation has
proved an effective proxy for wave breaking, which makes them well suited for
modeling coastal run-up. These features have led to mature, robust, and
efficient software for solving the SWE on realistic topographies, such as
GeoClaw \cite{leveque_tsunami_2011}, which combines a well-balanced and
positivity-preserving Riemann solver with patch-based adaptive mesh refinement.

However, the SWE have a constant phase velocity and thus completely neglect
dispersion. This limitation becomes important for tsunamis driven by
shorter-wavelength sources, such as those generated by submarine landslides or
volcanic events, and for the accurate prediction of leading-wave amplitudes
and arrival times over long propagation distances \cite{glimsdal2013dispersion}.
In these regimes, dispersive water wave models like the Serre-Green-Naghdi
(SGN) equations provide a more faithful description of the dynamics.
The cost of this improved fidelity is significant: dispersive systems involve
high-order and mixed space-time derivatives, whose discretization requires the
implicit inversion of differential operators at each time
step \cite{berger_towards_2021,berger_implicit_2023}. In addition, dispersive
models lack an inherent wave-breaking mechanism, which can lead to
unphysical results near the shore if not handled with care.

In recent years, hyperbolic relaxations of dispersive water wave models have been proposed as a 
way to circumvent the former difficulty. By introducing auxiliary variables and tunable relaxation 
parameters, the original mixed-derivative systems are replaced by first-order balance laws, which 
can be discretized with fully explicit, shock-capturing schemes while inheriting the favorable 
dispersion properties of the underlying dispersive models. 
Such reformulations of SGN-type models were introduced by Favrie and Gavrilyuk \cite{Favrie_2017} and have since been developed 
in several directions: mild-bottom variants \cite{escalante_efficient_2019,guermond2019robust},
full-bottom variants \cite{bassi_hyperbolic_2020,guermond_hyperbolic_2022},
and formulations with improved dispersion at moderate wavelengths \cite{duran2024discretization}.
The efficiency gain over direct SGN solvers has only recently begun to be assessed, with a regime-dependent payoff:
semi-implicit ImEx schemes yield substantial speedups at large relaxation parameters in 1D \cite{macca2025semi},
while fully explicit energy-conserving discretizations are favorable on coarse grids at moderate relaxation \cite{ranocha_ricchiuto_structure-preserving_2025}.
A complementary line of work addresses the latter difficulty by spatially coupling dispersive and non-dispersive models,
using the SWE to handle wave breaking near the 
shore \cite{tonelli2009hybrid,tissier2012new,bonneton_splitting_2011,kazolea_ricchiuto_2018,parisot_2024,galaz_2025}.
While this strategy is well established in practitioner codes, hyperbolic relaxations of dispersive models have, to our
knowledge, not yet been adopted in any of the major software packages used for tsunami modeling.
The present work brings the two together within GeoClaw's patch-based AMR framework.

\begin{figure}
    \centering
    \includegraphics[width=\linewidth]{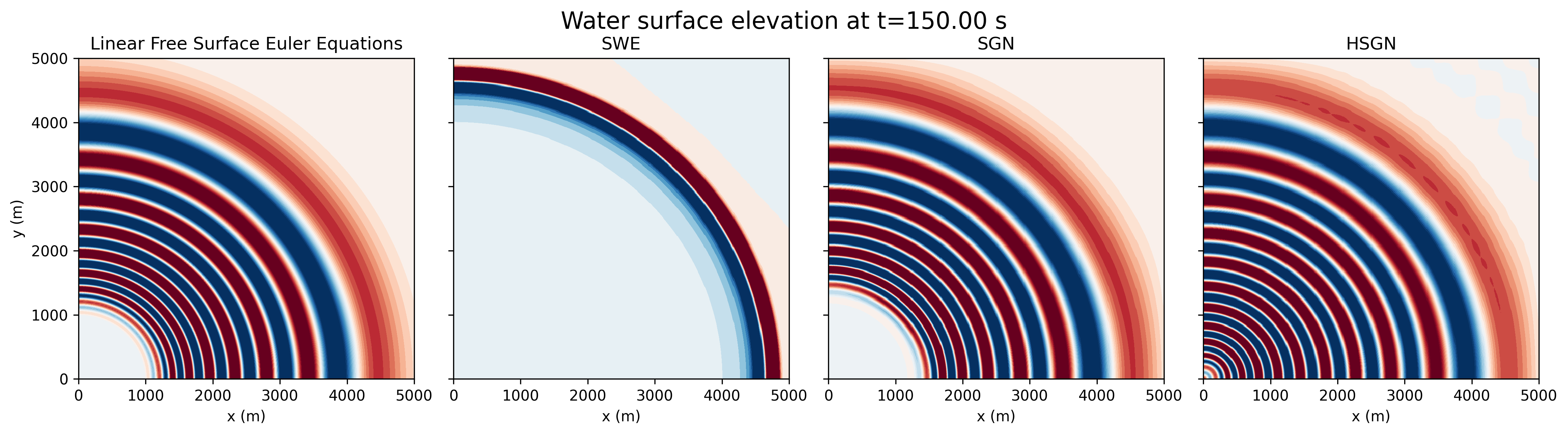}
    \caption{Free surface elevation at $t=150$ s for the linearized Euler
    equations (reference), SWE, SGN, and the hyperbolic approximation of SGN
    (HSGN) used in this work. The initial condition is a Gaussian hump
    centered at $(x,y)=(0,0)$ m over a water depth of $H=100$ m.}
    \label{fig:motivation_linear_Euler}
\end{figure}

As illustrated in Figure \ref{fig:motivation_linear_Euler} and developed
throughout this work, hyperbolic approximations of dispersive models can be
viewed as a middle ground between the SWE and the original dispersive systems,
both in terms of accuracy and computational cost. 
Moreover, they can be interpreted as approximations of the free surface Euler equations in their own right.
Combined with a transition
to the SWE close to the shore they offer
a single, fully explicit framework that captures dispersive propagation in
the deep ocean and robust wave breaking and run-up at the coast.

The goal of this paper is to develop, implement, and validate a hybrid
hyperbolic-dispersive/SWE solver which leverages the use
of patch-based adaptive mesh refinement.
The resulting algorithm:
\begin{itemize}
    \item is fully explicit, so that no large linear systems are solved at
    each time step;
    \item adapts both the model and the mesh in space and time, transitioning
    to the SWE where dispersion is irrelevant and wave breaking dominates;
    \item is implemented in GeoClaw and is thus immediately available to its wide
    user base, as an alternative to its current dispersive
    solvers \cite{berger_towards_2021,berger_implicit_2023};
    \item achieves a
    physical resolution comparable to that of the existing implicit SGN solver
    in GeoClaw \cite{berger_implicit_2023} at a reduced computational cost;
    \item discretizes a hyperbolic approximation of SGN, a mild-bottom variant,
    and is designed to be easily extendable to other hyperbolic approximations of dispersive
    water wave models in a straightforward manner; and
    \item is extensively validated against benchmarks and experimental data,
    as well as tested on a real tsunami scenario.
\end{itemize}
We also provide techniques for dealing with important issues inherent in such a solver, including
\begin{itemize}
    \item how and where to transition between dispersive and non-dispersive models;
    \item the choice of a relaxation parameter value that is physically adequate but still computationally advantageous;
    \item calculation of well-prepared initial data for the additional variables in the hyperbolic model;
    \item manufacturing of non-trivial steady state solutions, for verification of hyperbolic dispersive systems.
\end{itemize}

The rest of the manuscript is organized as follows.
Section \ref{sec:models} recalls the shallow water wave models we build on.
Section \ref{sec:hyperbolic models} introduces the hyperbolic approximations
and discusses their linear dispersion relation, well-prepared initial data,
and the transition to the SWE near the shore.
Section \ref{sec:methods} describes the discretization.
Verification of the solver is presented in Section \ref{sec:verification}, 
validation against experimental data and real tsunami events in Section \ref{sec:validation},
 and a cost and scaling comparison in Section \ref{sec: cost}.

 The code used to produce the results in this manuscript is available in
 the reproducibility repository \url{https://github.com/carlosmunozmoncayo/adaptive_explicit_HSGN}.
 The source code has also been permanently archived on Zenodo \cite{munozmoncayo_2026_hsgn_repro}.

\section{Shallow water models}
\label{sec:models}
\begin{figure}
\centering
\begin{minipage}[]{.49\textwidth}
  \centering
  \includegraphics[width=0.8\linewidth]{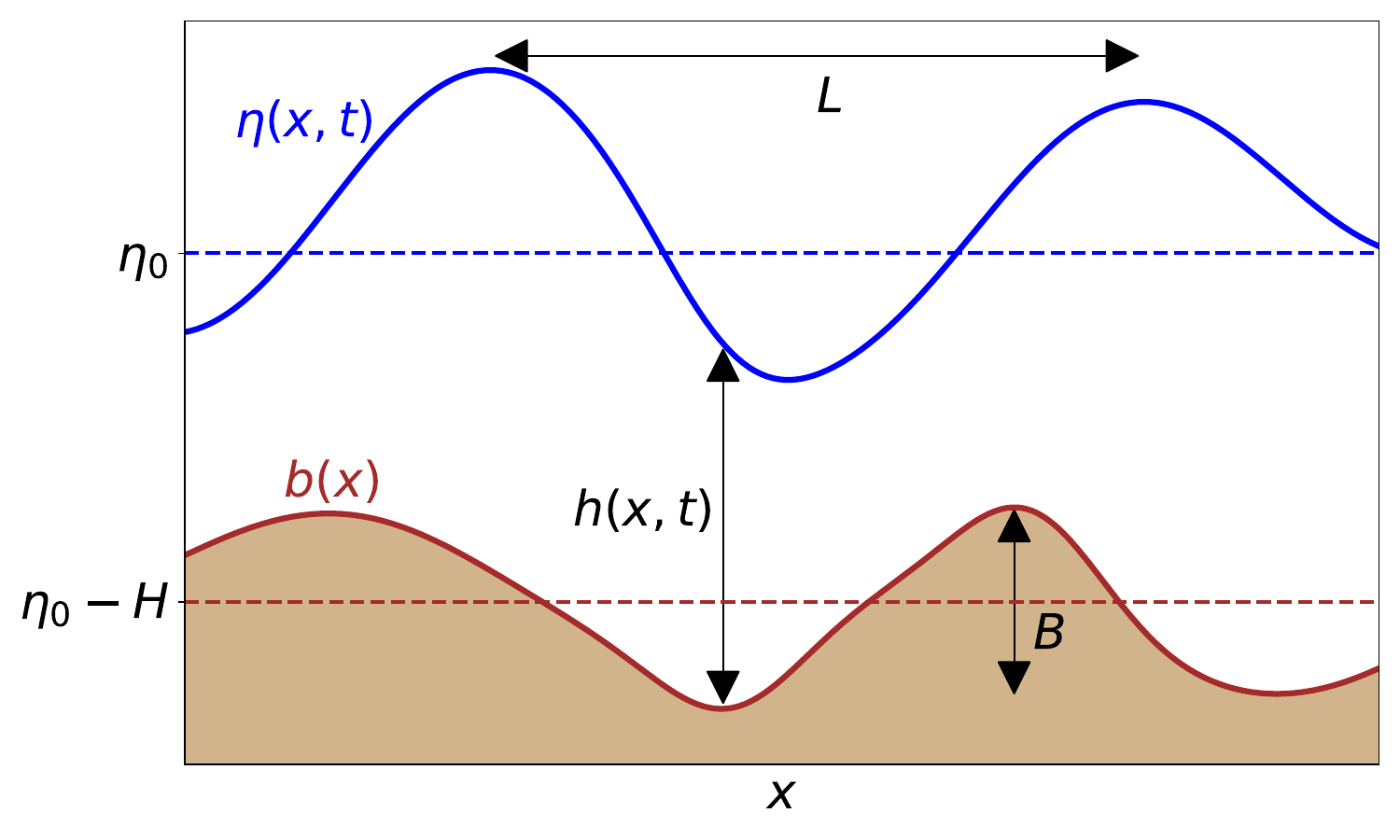}
  \captionof{figure}{Problem setting and notation}
  \label{fig:notation}
\end{minipage}%
\hfill
\begin{minipage}[]{.49\textwidth}
  \centering
  \vspace{0pt}
  \resizebox{\linewidth}{!}{%
\begin{tikzcd}[ampersand replacement=\&]
  \& {\text{Free surface Euler equations}} \& \\
  {\text{SWE}} \& {\text{SGN}} \& {\text{HSGN}} \\
  \& {\text{mbSGN}} \& {\text{mbHSGN}}
  \arrow["{\mathcal{O}(\mu^4)}", from=1-2, to=2-2]
  \arrow["{\mathcal{O}(\mu^2)}"', from=2-2, to=2-1]
  \arrow["{\mathcal{O}(1/c^2)}", from=2-2, to=2-3]
  \arrow["{\beta=\mathcal{O}(\mu^2)}", from=2-2, to=3-2]
  \arrow["{\mathcal{O}(1/c^2)}", from=3-2, to=3-3]
\end{tikzcd}}
\captionof{figure}{The models considered in this work and their formal asymptotic relationships,
where $\mu=H/L$, $\beta=B/H$ and $c^2$ is the hyperbolic relaxation parameter (see Section \ref{sec:hyperbolic models})}
  \label{fig:diagram models}
\end{minipage}
\end{figure}

Consider a body of water bounded above by its free surface elevation
$\eta(x,t)$ and below by the bottom topography $b(x)$, both measured relative to a constant sea level $\eta_0$ (which we set to zero without loss of generality)
as depicted in Figure \ref{fig:notation}. 
Under standard assumptions, 
the dynamics of the fluid are described by the incompressible Euler equations with a free surface 
\cite[Sec. 1.1.2]{lannes_water_2013}. Although these equations provide a complete description 
of the water wave dynamics, the fact that the upper boundary is unknown and time-dependent makes
their direct numerical simulation challenging, and prohibitively expensive for large-scale applications
such as tsunami modeling. A standard remedy is to derive simplified models that are more amenable 
to discretization while still capturing the essential features of the dynamics.
The most widely used simplifying assumption for tsunami modeling is that the characteristic 
wavelength
$\lambda$ is much larger than the characteristic water depth ${H}$,
i.e., $\mu = {H}/L \ll 1$.
Theoretical results \cite{Iguchi_2011} and experimental evidence \cite{Lay_2005} suggest that 
this assumption is valid for a broad class of earthquake-generated tsunamis, which is why the resulting 
shallow water models are widely used in practice. The starting point of our work consists of three such 
models: the Saint-Venant or shallow water equations (SWE), the Serre-Green-Naghdi (SGN) equations, and 
a mild-bottom approximation of SGN (mbSGN), all presented below in one spatial dimension; their
extension to two spatial dimensions is straightforward.

The classical Saint-Venant or shallow water equations (SWE) read 
\begin{subequations}
\label{eq:NSWEs}
    \begin{align}
& \partial_t h+\partial_x(h u)=0, \label{eq:NSWEs mass} \\
& \partial_t(h u)+\partial_x\left(h u^2+\frac{1}{2} g h^2\right)=-g h \partial_x b,\label{eq:NSWEs momentum}
\end{align}
\end{subequations}
where $h(x,t)=\max(0, \eta(x,t)-b(x))$ is the water depth, $u(x,t)$ is the depth-averaged horizontal fluid velocity, and $g$ is a gravitational acceleration constant. 
They are consistent with the Euler equations up to order $\bigo(\mu^2)$.
By considering a flat bottom and a constant depth $H$, perturbed by a small sinusoidal plane wave $\epsilon \sin(k x - \omega t)$,
one finds the linear dispersion relation $\omega(k)$ and the scaled phase velocity $C=\omega^2/(k^2 gH)$, which for the SWE is
$C_{\text{SWE}}=1$.  In other words, the SWE are non-dispersive, with all Fourier modes travelling at the same speed.
Real water waves are dispersive; the Euler equations with a free surface yield the normalized phase velocity
\begin{align}
    C_{\text{Euler}}=\frac{\omega_{\text{Euler}}^2(k)}{k^2 gH}
    = \frac{\mathrm{tanh}(k H)}{k H} 
    = 1 - \frac{(k H)^2}{3} + \bigo((k H)^4).
\end{align}
The SGN system approximates the Euler equations up to order $\bigo(\mu^4)$,
with wavenumber-dependent normalized phase velocity
$C_{\text{SGN}} =(1+\frac13(kH)^2)^{-1}=1 - \frac{(k H)^2}{3} + \bigo((k H)^4)$.  

A first-order reformulation of SGN
was introduced in \cite{fernandez2018hierarchy}:
\begin{subequations}
\label{eq: first order SGN}
\begin{align}
& \partial_t h+\partial_x(h u)=0 ,\\
& \partial_t(h u)+\partial_x\left(h u^2+\frac{1}{2} g h^2+h p\right)=-\left(g h+p_b\right) \partial_x b ,\\
& \partial_t(h w)+\partial_x(h {u w})=p_b, \\
& \partial_t(h \sigma)+\partial_x(h u \sigma)=-6 p_b+12 p, \\
& \sigma=-h \partial_x u, \\
& w-\frac{\sigma}{2}-u \partial_x b=0.
\end{align}
\end{subequations}
Here $w$ is the vertical depth-averaged velocity, $\sigma$
is an auxiliary variable encoding the depth-integrated horizontal divergence,
$p$ is the depth-averaged non-hydrostatic pressure, and $p_b$ is the non-hydrostatic pressure evaluated at the bottom.
The advantage of this formulation is that all time derivatives act on conserved quantities,
which makes it a natural starting point for the hyperbolic relaxation discussed in Section \ref{sec:hyperbolic models}.
A simpler first-order formulation that avoids the auxiliary variable $\sigma$
was proposed in \cite{bristeau_energy-consistent_2015}; it reads
\begin{subequations}
\label{eq:Bristeau}
\begin{align}
& \partial_t h+\partial_x(h u)=0, \\
& \partial_t(h u)+\partial_x\left(h u^2+\frac{1}{2} g h^2+h p\right)=-\left(g h+\frac32 p\right) \partial_x b, \\
& \partial_t(h w)+\partial_x(u h w)=\frac32 p, \\
& \partial_x u+\frac{w+u \partial_x b}{h / 2}=0. \label{eq: closed-form w mbSGN}
\end{align}
\end{subequations}
System \eqref{eq:Bristeau} is derived under the additional assumption of mild bottom variations, i.e., $\beta = \mathcal{O}(\mu^2)$ \cite[Sec. 3.3]{guermond_hyperbolic_2022},
where $\beta=B/H$ is the characteristic bottom amplitude. Thus, we refer to it as the mild-bottom Serre-Green-Naghdi (mbSGN) system.
As expected, mbSGN and SGN coincide when the bottom is flat and, hence, they share the same dispersion relation.
The structural simplicity of mbSGN, in particular the absence of $\sigma$,
translates into a correspondingly simpler hyperbolic relaxation.

\section{Dispersive hyperbolic models}
\label{sec:hyperbolic models}
This work is concerned with the discretization of hyperbolic
approximations of the two dispersive models introduced above:
a hyperbolic approximation of mbSGN, proposed in \cite{escalante_efficient_2019},
and a hyperbolic approximation of SGN, proposed in \cite{bassi_hyperbolic_2020}. 
We refer to them as mbHSGN and HSGN, respectively. In one spatial dimension, mbHSGN reads
\begin{subequations}
\label{eq:mbHSGN}
\begin{align}
& \partial_t h+\partial_x(h u)=0, \\
& \partial_t(h u)+\partial_x\left(h u^2+\frac{1}{2} g h^2+h p\right)=-\left(g h+\frac32 p\right) \partial_x b, \\
& \partial_t(h w)+\partial_x(u h w)=\frac32 p, \\
& {\partial_t(h p)+\partial_x(u h p)+h c^2}\left(\partial_x u+\frac{w-u \partial_x b}{h / 2}\right)=0,  \label{eq: constraint w 1D}
\end{align}
\end{subequations}
while HSGN reads
 \begin{subequations}
\label{eq: HSGN}
\begin{align}
& \partial_t h+\partial_x(h u)=0, \\
& \partial_t(h u)+\partial_x\left(h u^2+\frac{1}{2} g h^2+h p\right)=-\left(g h+p_b\right) \partial_x b ,\\
& \partial_t(h w)+\partial_x(h u w)=p_b \\
& \partial_t(h \sigma)+\partial_x(h u \sigma)=-6 p_b+12 p, \\
& \partial_t(h p)+\partial_x\left[h u\left(p+c^2\right)\right]-c^2 u \partial_x h=-c^2 \sigma, \\
& \partial_t\left(h p_b\right)+\partial_x\left(h u p_b\right)-6 c^2 u \partial_x b=-6 c^2\left(w-\frac{\sigma}{2}\right).
\end{align}
\end{subequations}
These systems are hyperbolic singular perturbations of mbSGN and SGN in the sense that, as $c^2\to \infty$,
the solutions of \eqref{eq:mbHSGN} and \eqref{eq: HSGN} are expected to converge to those of mbSGN and SGN, respectively.
A direct computation shows that these systems are hyperbolic as long as  $g h+c^2\geq |p|$.

It can be shown that the eigenvalues of the flux Jacobian of mbHSGN and HSGN are bounded by
\begin{align}
    \label{eq: eigenvalue_bound}
    \lambda_{\max/ \min} =  u \pm \sqrt{g h +  p + c^2}.
\end{align}
This implies that large values of $c^2$ will lead to stiffness and require the use of implicit time discretization.
However, as discussed in Section \ref{sec:LDR} below, a moderate value of $c^2$ yields a sufficiently
accurate approximation for our purposes, which motivates the use of an explicit discretization.
Following \cite{escalante_efficient_2019} and \cite{bassi_hyperbolic_2020},
we introduce the dimensionless relaxation parameter as $\alpha = c/\sqrt{g H}$, which controls the 
tradeoff between accuracy and stiffness.

\subsection{Linear dispersion relation}
\label{sec:LDR}
\begin{figure}
    \centering
    \includegraphics[width=\linewidth]{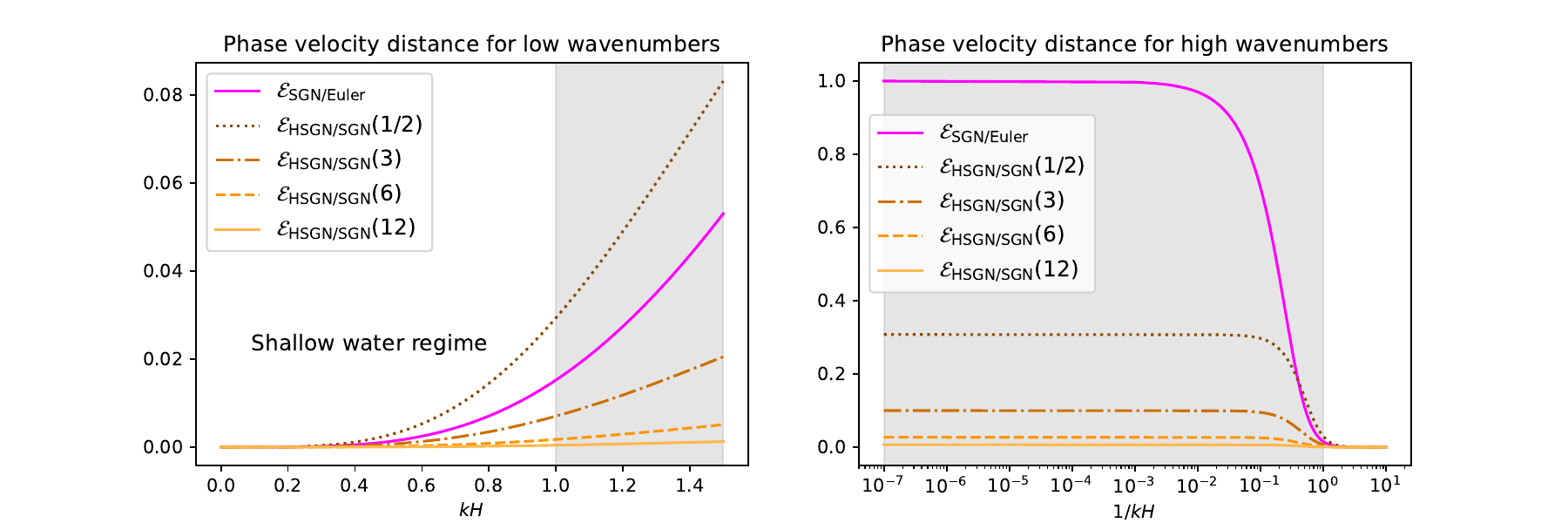}
    \caption{Relative distance of the phase velocity of HSGN with respect to SGN \eqref{eq: distance phase velocity SGN} for different values of $\alpha=c/\sqrt{gH}$ (orange lines).
    The relative distance of the phase velocity of SGN with respect to the Euler equations \eqref{eq: distance phase velocity SGN Euler} is also plotted for comparison (magenta line).
    The shallow water regime corresponds to the white region of the plot, while the moderate and deep water regimes correspond to the light gray regions.}
    \label{fig:LDR}
\end{figure}

These hyperbolic models yield a remarkably accurate linear dispersion
relation, even for moderate values of the dimensionless relaxation
parameter $\alpha$.
A key observation in  \cite[Sec. 2.3]{escalante_efficient_2019} and \cite[Sec. 2.3]{bassi_hyperbolic_2020}
is that
the error in the scaled phase velocity of the hyperbolic models with respect to that of
the free surface Euler equations
remains below a few percent even as we depart from the shallow water regime.
Conceptually, one introduces a first modeling error by replacing the Euler equations with the SGN system,
and a second modeling error by replacing SGN with its hyperbolic approximation.
To quantify these errors,
let us denote them in relative terms by 
\begin{subequations}
\begin{align}
    \label{eq: distance phase velocity SGN}
    \mathcal{E}_{\text{HSGN/SGN}}(\alpha) = \frac{|C_{\text{HSGN}}^\alpha-C_{\text{SGN}}|}{|C_{\text{SGN}}|}, \\
    \mathcal{E}_{\text{SGN/Euler}} = \frac{|C_{\text{SGN}}-C_{\text{Euler}}|}{|C_{\text{Euler}}|}. \label{eq: distance phase velocity SGN Euler}
\end{align}
\end{subequations}
Then, as illustrated in Figure \ref{fig:LDR}, the relaxation parameter can be chosen such
that $\mathcal{E}_{\text{HSGN/SGN}}(\alpha) < \mathcal{E}_{\text{SGN/Euler}}$ for every wave number $kH$.
In fact, this can be achieved for a value of $\alpha$ as small as $\alpha=3$,
which in turn, reduces the artificial stiffness of the system and thus
allows for an efficient explicit discretization.

Finally, we note that the second (hyperbolization) error $\mathcal{E}_{\text{HSGN/SGN}}(\alpha)$ 
is largest at shorter wavelengths,
where models such as SGN or SWE are not expected to provide an accurate
description of the water wave dynamics.  
Of course, in practical applications we may deal with some waves outside the long wavelength (shallow water) regime,
and it is desirable to work with models that remain well-behaved in such scenarios \cite[Ch. 5.2]{lannes_water_2013}.
We provide some numerical observations regarding the behavior of mbHSGN and HSGN
in the short wavelength regime in Section \ref{sec:validation}.  For a recent contribution to the improvement of the 
dispersion relation of hyperbolic models in that regime, see \cite{duran2024discretization}.
\subsection{Well-prepared initial data}
\label{sec:well-prepared}
It is well-known that for singularly perturbed problems,
the choice of initial data can have a significant impact on the accuracy with which the solution of the
relaxation system approximates that of the original one \cite{giesselmann2026convergence}.
In particular, it is desirable to
initialize the additional variables in a manner that is consistent with the 
constraints that would be satisfied by the solution of the original dispersive system.
We refer to such initial data as well-prepared, and to the opposite as ill-prepared.

We illustrate how this can be achieved for mbHSGN. 
Let us assume that we have expressions for $h$ and $u$.
Then, we immediately obtain $w$ from the constraint \eqref{eq: closed-form w mbSGN}, while
for $p$ we get the relation 
\begin{subequations}\label{p-initial}
\begin{align}
    \label{eq: constraint p 1D}
p
=-\frac23 \partial_x(huw) + \frac23\partial_x(hu) w + \frac{h}{3}  \left(-\partial_x(hu) \partial_x u+\partial_x\partial_t u+\partial_t u \partial_x b \right),
\end{align}
where $\partial_t u$ can be expressed as
\begin{align}
    \label{eq: ut 1D}
    \partial_t u = \frac{1}{h}\left[-\partial_x(hu^2+\frac12 g h^2) - \partial_x h p-h\partial_x p -(gh+\frac32 p)\partial_x b+u\partial_x(h u)\right].
\end{align}
\end{subequations}
After discretizing in space, \eqref{p-initial} becomes a system of linear algebraic equations
that we can solve to obtain well-prepared initial data for $p$.
Alternatively, if available, a solver for SGN can be evolved for a few small time steps and $\partial_t u$ can
be approximated using e.g., a finite difference in time, which allows us to compute $p$
from \eqref{eq: constraint p 1D} without the need to solve a linear system.
The latter is the approach we take in this work whenever well-prepared initial data is required,
unless explicit formulas are available, e.g., for solitary waves and steady states.

\subsection{Wave breaking and transition to the shallow water equations}
\label{sec: wave breaking}
As waves approach the shore, shoaling takes place: waves steepen and eventually break,
forming bores and turbulent flows in which energy is dissipated.
Robustly modeling wave breaking is a key requirement for tsunami simulation,
since it directly affects coastal run-up and the resulting hazard.
In the SWE, wave breaking and the associated energy dissipation are modelled naturally
through shock formation.
Dispersive models such as SGN, in contrast, must be augmented with an explicit breaking criterion
(based, for example, on local wave steepness or energy dissipation)
together with a dissipative mechanism, typically artificial viscosity or a local transition to the SWE.
We refer to \cite{kazolea_ricchiuto_2018} for a comprehensive review.

The approach followed in this work is to transition to the SWE close to the shore, i.e., where the water depth of the ocean at
rest is smaller than a given threshold, and to take into account dispersive effects everywhere else, as done in e.g., \cite{berger_towards_2021,berger_implicit_2023}.
This is achieved by allowing the relaxation parameter to vary in space and time in such a way that it transitions from a moderate
value in the deep ocean to zero close to the shore. For instance, we choose
\begin{align}
\label{eq: space dependent relaxation parameter}
    c^2=\alpha g H(x) \phi_T(x),
\end{align}
where $\phi_T(x)$ is a transition function valued between 0 and 1, to be specified  in Equation \eqref{eq:transition_function} below
and $H(x)$ is the depth of the ocean at rest.
The rationale behind this approach is that,
as was noted in \cite{escalante_efficient_2019}, not only the original dispersive systems are recovered in the limit $c^2\to \infty$, but also the SWE are recovered
as $c^2\to 0$ (as long as the non-hydrostatic pressure also vanishes).
A benefit of this hybrid approach is that we can take advantage of the robustness of a well-tested and widely used solver for the 
SWE \cite{george2008augmented} to handle wave breaking and coastal run-up, while still incorporating dispersive effects in the rest of the domain.
Moreover, while on small-scale laboratory experiments, a fixed reference depth (and thus a fixed value of $c^2$) can be chosen to achieve a good approximation of the wave dynamics,
in large-scale applications such as tsunami modeling, the water depth can vary by several orders of magnitude, and thus a spatially varying $c^2$ 
is mandatory to prevent prohibitive stiffness close to the shore.

\section{Discretization}
\label{sec:methods}
In two spatial dimensions, mbHSGN \eqref{eq:mbHSGN} and HSGN \eqref{eq: HSGN} can both be written in the form
\begin{equation}
    \label{eq: compact 2D}
    \partial_t \bfq + \nabla \cdot \bfF(\bfq, \bfx) +
    \bfA(\bfq,\bfx) \cdot \nabla \bfq +
     \bfB(\bfq,\bfx) \nabla b = \bfS(\bfq,\bfx).
\end{equation}
The terms on the left-hand side, which include spatial derivatives, represent a
non-conservative hyperbolic system, while those on the right (the \emph{source terms}) are purely algebraic
and are responsible for dispersion.
We treat the hyperbolic part and the source separately via operator splitting (Lie-Trotter or Strang),
on (possibly mapped) logically rectangular grids with the Berger-Oliger-Colella patch-based AMR algorithm \cite{berger_adaptive_1998}.
AMR is essential here, as realistic geophysical applications span many spatial and temporal scales.

For the hyperbolic part, we use a wave-propagation formulation of
Godunov-type methods,
following the  framework used for the SWE in GeoClaw \cite{leveque_tsunami_2011}.
A central component of the method is the use of a normal Riemann solver at cell interfaces, and
a transverse Riemann solver at cell corners, which allows for a more stable discretization with a larger time step (i.e. a CFL number closer to 1 in two dimensions)
with second-order accuracy in space and time \cite{LEVEQUE1997}.
At each interface, the system is reduced to a 1D Riemann problem in the normal direction with the tangential velocity passively advected,
and slope limiting (minmod and MC herein) is applied to the resulting waves.

We use a Riemann solver based on the $f$-wave 
decomposition \cite{bale_wave_2003,george2008augmented}, which allows us to handle the 
space-dependent flux functions and non-conservative products
naturally. 
At the interface between cells $i-1,j$ and $i,j$, the total fluctuation
\begin{subequations}
    \label{eq:total fluctuation}
    \begin{align}
    \mathcal{A}\Delta \bfQ_{i-1/2,j} =& \bfF^x(\bfQ_{i,j},\bfx_{i,j})-\bfF^x(\bfQ_{i-1,j},\bfx_{i-1,j})\\
    &+\int_{\bfx_{i-1,j}}^{\bfx_{i,j}} \bfA^x(\bar{\bfQ}(s),s) \bar{\bfQ}'(s) + \bfB^x(\bar{\bfQ}(s),s) \bar{b}'(s) ds,
\end{align}
\end{subequations}
is decomposed into $f$-waves proportional to the eigenvectors of $\partial \bfF^x/\partial \bfq + \bfA^x$,
where $\bfF^x$, $\bfA^x$, and $\bfB^x$ are the flux, non-conservative product, and bathymetry-dependent terms in the $x$-direction,
respectively.
Here $\bfQ_{i,j}$ is the cell average of the conserved variables and $\bfx_{i,j}$ is the cell center.
The function $\bar{\bfQ}(s)$ is a Lipschitz continuous path in state space
connecting the left and right states $\bar{\bfQ}_{i-1,j}$ and $\bar{\bfQ}_{i,j}$ \cite{leveque2011well, pares_2006}.
We simply approximate the integral in \eqref{eq:total fluctuation} with a one-point quadrature rule using Roe-type averages.
Specifically, the non-conservative product integral is approximated by
$$
\bfA^x(\hat{\bfQ}_{i-1/2,j},\hat{\bfx}_{i-1/2,j})(\bfQ_i-\bfQ_{i-1}).
$$
The interface state $\hat{\bf Q}_{i-1/2,j}$ uses arithmetic means
$\hat{h}_{i-1/2,j} = \tfrac12(h_{i,j}+h_{i-1,j})$ and $\hat{c}^2_{i-1/2,j} = \tfrac12(c^2_{i,j}+c^2_{i-1,j})$
for the water height and relaxation parameter, and depth-weighted Roe-type averages
\begin{align}
    \hat{m}_{i-1/2,j} = \frac{\sqrt{h_{i,j}}\, m_{i,j} + \sqrt{h_{i-1,j}}\, m_{i-1,j}}{\sqrt{h_{i,j}} + \sqrt{h_{i-1,j}}}
\end{align}
for the remaining primitive variables ($\bfu, w, p$ for mbHSGN, with $\sigma$ and $p_b$ added for HSGN).

Regarding the bathymetry-dependent terms, for mbHSGN (and analogously for HSGN) we use the following approximations
\begin{align}
    \int_{\barx_{i-1}}^{\barx_{i}} \left(g\barh+\frac32 \barp\right) \barb' ds &\approx \left(g \hat{h} + \frac32 \psi(b_{i},b_{i-1}) \hat{p}\right) (b_{i}-b_{i-1}),\\
    \int_{\barx_{i-1}}^{\barx_{i}} -\barc^2 \baru \left(\barh'+2 \barb'\right) ds &\approx -\hat{c}^2 \hat{u} \left(\eta_{i}-\eta_{i-1} + \psi(b_{i},b_{i-1}) (b_{i}-b_{i-1})\right),
\end{align} 
where $\psi$ is a bathymetry sharpness sensor defined as 
\begin{align}
    \psi(b_{i},b_{i-1}) = \begin{cases}
        1, & \text{if } \xi < \xi_0,\\
        -3((\xi-\xi_0)/\xi_1)^2 + 2((\xi-\xi_0)/\xi_1)^3+1, & \text{if } \xi_0 \leq \xi < \xi_0+\xi_1,\\
        0, & \text{otherwise},
    \end{cases}
\end{align}
with $\xi = 2|b_i-b_{i-1}|/|b_i+b_{i-1}|$, and the user-defined parameters $\xi_0,\,\xi_1>0$, which control the sensitivity of the sensor 
to relative bathymetry variations.
The sensor suppresses oscillations near steep bathymetry gradients, 
which the hyperbolic relaxation tends to amplify (see e.g.\ \cite[Sec. 4.2]{bassi_hyperbolic_2020} and supplementary material),
while recovering a consistent discretization on smooth bathymetries and fine grids.
The $f$-waves are the eigenvectors of the system linearized about the Roe-type averages,
with Einfeldt speeds
$$
s_{\min}= \min(\lambda_{\min}(\hat{\bfQ}_{i-1/2,j}), \lambda_{\min}({\bfQ}_{i-1,j})),\,
s_{\max}= \max(\lambda_{\max}(\hat{\bfQ}_{i-1/2,j}), \lambda_{\max}({\bfQ}_{i,j})),
$$
for the genuinely nonlinear waves, and linearly degenerate eigenvalues evaluated at $\hat{\bfQ}_{i-1/2,j}$.
The resulting discretization is exactly well-balanced for the ocean at rest. 
One may proceed similarly in the $y$-direction at the interface between cells $i,j-1$ and $i,j$.
A transverse Riemann solver is constructed analogously from the known eigenstructure of the system.

For some problems the use of mapped grids is convenient, as it allows us to handle immersed boundaries aligned with the grid
and, more importantly, to work with latitude-longitude grids for large-scale applications (see \cite[Ch. 8.2]{george2006finite}).
The extension of the method described above to mapped grids follows the standard approach presented in \cite[Ch. 23]{LeVeque_2002},
where rotated data is used to solve the Riemann problem at cell interfaces,
and the resulting fluxes and wave speeds are scaled by the aspect ratios between computational and physical space.

For the source-term step, the ODEs $\partial_t \bfq=\bfS(\bfq,\bfx)$ at each cell are linear in the primitive variables ($\bfu, w, p, \sigma, p_b$),
and can be solved exactly via a small matrix exponential ($2\times 2$ for mbHSGN, $4\times 4$ for HSGN).
However, these systems are only neutrally stable, and rounding errors can lead to amplification of unstable modes 
when using the exact solution in long-time simulations.
We instead use the two-stage, second-order, L-stable SDIRK(2,2) method \cite[Sec 4.1.2]{kennedy2016diagonally},
which we found to balance accuracy and robustness well across our test cases.

Regarding the transition to the SWE, given a spatial resolution $\Delta x$ and a number of 
neighbors $n$, we consider the following transition function
\begin{align}
    \label{eq:transition_function}
    \phi_T(\bfx) =
    \begin{cases}
        1, &
        \text{if }
            \min_{|\bfx-\xi|\leq n\Delta x}
            \left((\eta_0-b(\xi))^+\right) > h^*, \\
        0, & \text{otherwise},
    \end{cases}
\end{align}
where $h^*$ is a threshold depth below which we solve the SWE, and $\eta_0$ is the initial free surface elevation.
For all the test cases considered below, we choose $n=1$, e.g., a neighborhood of one cell around the wet-dry front is always
transitioned to the SWE, which allows us to preserve the water depth positivity for very coarse grids.
The choice of the transition threshold will depend on the test cases considered.
While small oscillations can be observed at the interface between models,
these do not seem to affect the robustness of the method and
they appear to have much less significant impact on the solution
than the reflections produced by, e.g., the bathymetry.
Designing a more sophisticated transition, possibly informed by global energy preservation, is left for future work.
Progress in this direction has been made recently in the context of coupling 
dispersive and non-dispersive water wave models, e.g., in \cite{parisot_2024,galaz_2025}.

The time step follows the standard CFL condition with global maximum wave speeds, with subcycling in time across AMR levels.
Unless otherwise specified, we use a Courant number between 0.9 and 1, sensor parameters $\xi_0=0.01$ and $\xi_1=0.05$, 
and non-dimensional relaxation parameter $\alpha=3$ (except in Sections~\ref{sec: nontrivial steady} and \ref{sec: cost}).

\section{Verification of the scheme and manufactured solutions}
\label{sec:verification}
In this section we verify the discretization described in Section~\ref{sec:methods} 
against known solutions and against high-resolution reference solutions of the original dispersive systems.
The tests progress from a 1D Gaussian hump that isolates the effect of well-prepared initial data and the hyperbolization error,
through solitary waves and manufactured non-trivial steady states that probe convergence rates, to 2D radial and oblique tests
that exercise the full scheme with AMR. 
Throughout, reference solutions of the SGN system \eqref{eq: first order SGN} are computed
using the solver proposed by Berger and LeVeque \cite{berger_implicit_2023}\footnote{In order to isolate the effect of the
 hyperbolic approximation from that of the improved dispersion relation,
 for the solution of SGN
 we use the parameter $\alpha=1$ in the notation of \cite{berger_implicit_2023},
 not to be confused with the non-dimensional hyperbolic relaxation parameter in this work.}.

\subsection{Well-prepared initial data and hyperbolization error}
\label{sec: hump flat bottom 1D}
We consider the scenario of a Gaussian hump in the surface elevation over flat bathymetry with
zero initial velocity:
\begin{align}
  \label{eq:initial Gaussian hump}
    h(x,0) = H + A \exp\left(-\frac{(x-x_0)^2}{\sigma^2}\right),\quad u(x,0)=0 \si{\meter}/\si{\second},\quad b(x)=-H.
\end{align}
We take $H=100 \si{\meter}, A=5 \si{\meter}, x_0=0 \si{\meter}, \sigma=100 \si{\meter}$, and the domain is $[-10^4 \si{\meter}, 10^4 \si{\meter}]$.  
The solution is evolved until $t=75 \si{\second}$ with $3\times 10^4$ cells.
Since HSGN is equivalent to mbHSGN over a flat bottom, we only consider mbHSGN for this scenario.
 
\begin{figure}
  \centering
    \includegraphics[width=\linewidth]{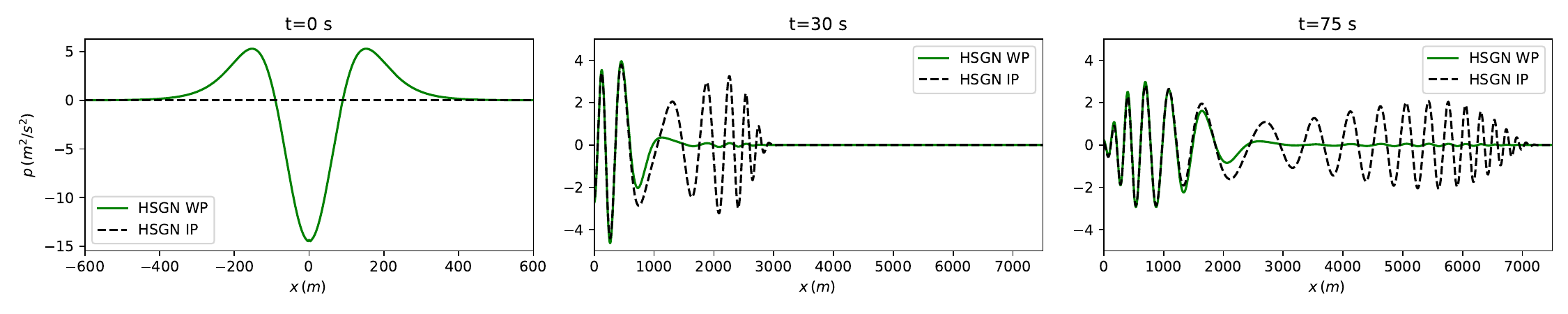}
    \caption{Evolution of non-hydrostatic pressure $p$ for HSGN with well-prepared (green solid line) and ill-prepared (black dashed line) initial data.  Note the different axis ranges between panels.}
    \label{fig: Gaussian hump well-prepared vs not p}
\end{figure}

We first use this test to illustrate the effect of well-prepared versus ill-prepared initial data. 
It is clear from \eqref{eq: constraint w 1D} that the vertical velocity $w$ should be identically zero at $t=0$. 
Determining an appropriate well-prepared condition for the non-hydrostatic pressure $p$ is more involved.
Proceeding as outlined in Section \ref{sec:well-prepared} (i.e. evolving SGN for three time steps of size $\Delta t=10^{-5} \si{\second}$ and 
approximating $\partial_t(hw)(x,0)$ through a one-sided finite difference), 
we obtain a non-trivial initial profile for $p$ that is shown in the left panel of Figure \ref{fig: Gaussian hump well-prepared vs not p}.
In the center and right panels of the same figure, we show the evolution of $p$ for HSGN with well- and ill-prepared ($p=w\equiv0$) initial data (we set $w=0$ in both cases).
We see that the slow-moving part of the solution (near the origin) is nearly identical.  However,
the ill-prepared initial condition generates a train of faster-moving unphysical waves in $p$.
Their speed is approximately equal to the spurious phase velocity $C^+_p$ (see \cite[Eq. 15]{escalante_efficient_2019}),
which arises from the hyperbolic approximation and grows with the relaxation parameter $c^2$.

\begin{figure}
  \centering
    \includegraphics[width=\linewidth]{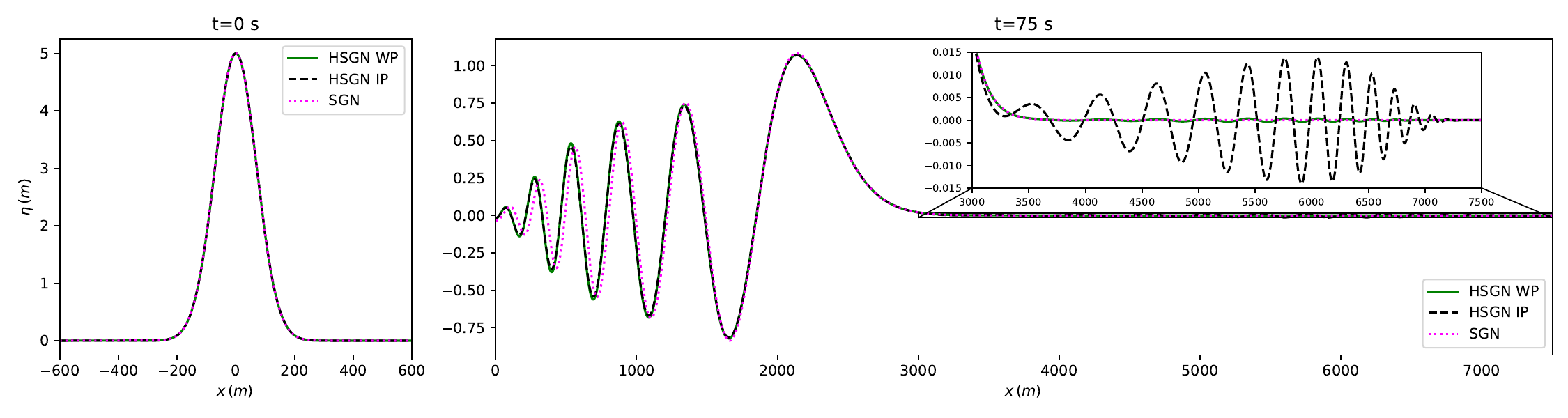}
    \caption{Evolution of surface elevation $\eta$ for HSGN with well-prepared (green solid line) and ill-prepared (black dashed line) initial data.
    The solution of SGN is shown in magenta dotted line for reference.}  
    \label{fig: Gaussian hump well-prepared vs not eta}
\end{figure}

In Figure \ref{fig: Gaussian hump well-prepared vs not eta} we show the evolution of the surface elevation $\eta$ for both cases
and compare it with the solution of SGN.
A careful inspection reveals the presence of the fast-moving waves, as well as a very small reduction in
accuracy of the slower-moving part of the solution; indeed, the difference between the well- and ill-prepared
solutions is much smaller than the difference between either of them and the SGN solution.
This can be explained by the fact that the solution is very close to a hydrostatic equilibrium, and thus the non-hydrostatic pressure is small compared to the hydrostatic one,
which is well captured by the initial condition in both cases.

\begin{figure}
    \centering
    \includegraphics[width=\linewidth]{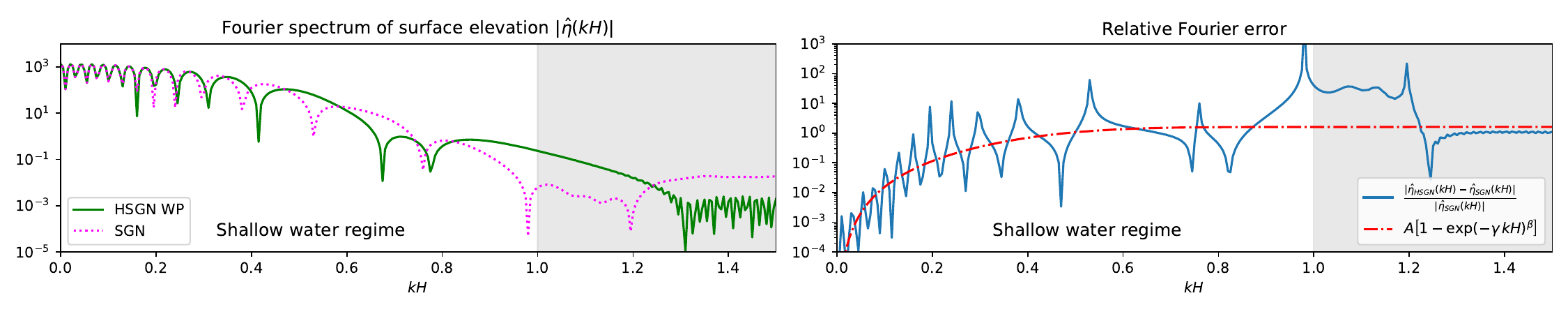}
    \caption{Spectral content of the solution at $t=25 \si{\second}$. Left panel: Fourier amplitude of the surface elevation $|\hat{\eta}|$
    for HSGN and SGN. 
    Right panel: relative difference between the models in Fourier space.  The dashed line is a fitted exponential ($A \approx 1.6, \gamma \approx 2.1, \beta \approx 2.9$).}
    \label{fig: Fourier analysis}
\end{figure}

Inspection of the surface elevation profile also reveals that
the hyperbolization error is larger at higher frequencies. 
This can be seen in Figure \ref{fig: Fourier analysis}, where we plot (in the left panel)
the absolute value of the Fourier transform of $\eta$ for both the SGN and HSGN solutions.
The relative difference between them is plotted in the right panel, and roughly follows an exponential distribution.

\subsection{Propagation of solitary wave solutions}
Next we validate the accuracy of the numerical scheme by simulating the propagation
of a solitary wave for HSGN and mbHSGN. 
The initial profiles, dependent on a single similarity
variable, are generated by solving a traveling wave ODE system (see \cite[Sec 5.1]{escalante_efficient_2019} and \cite[Sec 4.1]{bassi_hyperbolic_2020}) with
an 8th order Runge-Kutta method.
As initial condition we use a perturbation in the direction of the unstable manifold of the equilibrium point of the ODE system,
which corresponds to the ocean at rest.
We choose a wave of amplitude $a=0.2 \si{\meter}$ over a background depth of $H=1 \si{\meter}$,
and evolve the solution a full period over a periodic domain chosen large enough to avoid interactions with the boundaries.
The results of the convergence test, where the expected rate of convergence is attained,
 are shown in Table \ref{tab:exact_solitary_convergence}.

\begin{table}[]
\centering
\begin{tabular}{|c|c|c|c|c|c|c|}
\hline
 & {N}  & 64 & 128 & 256 & 512 & 1024 \\ \hline
\multirow{2}{*}{{mbHSGN}} 
 & {$L^\infty$ Error $h$} 
 & 9.38e-03 & 2.68e-03 & 6.93e-04 & 1.90e-04 & 6.25e-05 \\ \cline{2-7}
 & {Rate} 
 & - & 1.81 & 1.95 & 1.87 & 1.60 \\ \hline
\multirow{2}{*}{{HSGN}} 
 & {$L^\infty$ Error $h$} 
 & 2.14e-02 & 5.96e-03 & 1.46e-03 & 3.31e-04 & 7.87e-05 \\ \cline{2-7}
 & {Rate} 
 & - & 1.84 & 2.03 & 2.14 & 2.07 \\ \hline
\end{tabular}
\caption{Convergence test for a solitary wave over a flat bottom}
\label{tab:exact_solitary_convergence}
\end{table}

\subsection{Non-trivial steady-state solutions}
\label{sec: nontrivial steady}
To verify our numerical scheme in a scenario where the topography
is non-flat we propose a family of non-trivial steady-states for the 
mbHSGN system \eqref{eq:mbHSGN} that can be obtained by prescribing the vertical non-hydrostatic
velocity $w$, as is done in \cite{bristeau_energy-consistent_2015} for the mbSGN system.
\begin{proposition}
    Given a constant momentum $hu=q$, a hyperbolic 
    relaxation parameter $c^2>0$, a function $w\in \mathcal{C}_0^2(\real)$, and boundary conditions $h(x_0)$ and $b(x_0)$,
    the mbHSGN system \eqref{eq:mbHSGN} has an analytical steady-state solution satisfying 
\begin{align}
  \label{eq: steady-state EDC}
  p = \frac{3 q}{2} w',\quad
  h' = \frac{-h p' - \left(g h + \frac32 p\right) \left(\frac{h \, p'}{2\,c^2} + \frac{w h}{q}\right)}
  {\frac12 g h - \frac{q^2}{h^2} + \left(1 - \frac34 \right) p},\quad
  b' = -\frac12 h'+ \frac{h\,w}{q}+\frac{h\, p'}{2 c^2}.
        \end{align}
\end{proposition}
Note that the ODE system for the mbSGN steady state
\cite[eqn. (73)]{bristeau_energy-consistent_2015}
is formally recovered in the relaxed limit $c^2\to \infty$. 

\begin{figure}
    \centering
        \centering
    \includegraphics[width=\linewidth]{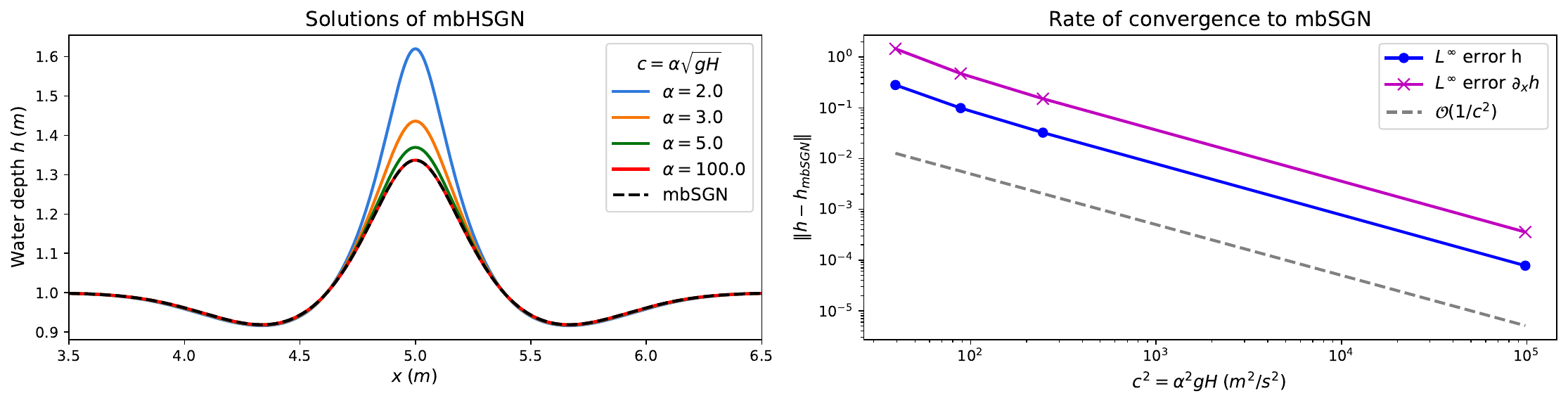}
    \caption{Convergence of the steady-state solutions of mbHSGN to the steady-state solution of the mbSGN system 
    as the relaxation parameter $c^2$ increases. Left panel: water height $h$ obtained by solving \eqref{eq: steady-state EDC} 
    using an 8th order Runge-Kutta method for different values of $c^2$. Right panel: rate of convergence in $C^1$ norm.}
    \label{fig: steady state convergence mbHSGN}    
\end{figure}

In Figure \ref{fig: steady state convergence mbHSGN}, we plot the solution of
\eqref{eq: steady-state EDC} for the particular choice of
\begin{align}
    w(x)= 2a_3(x - a_1)e^{-a_2(x - a_1)^2},
\end{align}
with $a_1=5 \si{\meter}, a_2=\frac{17}{5} \si{\meter}^{-2}$, and $a_3=\frac{3}{2} \si{\second}^{-1}$.
By varying the value of $c^2$, we see that 
the steady-state solutions of the mbHSGN system \eqref{eq:mbHSGN} seem to converge
at a rate of $\mathcal{O}(1/c^2)$ to the steady-state solution of the mbSGN system.
This is consistent with was obtained for solitary wave solutions to HSGN over a flat bottom
in \cite[Sec 4.1]{bassi_hyperbolic_2020}.
In the recent work \cite{giesselmann2026convergence}, a 
linear rate of convergence in the relaxation parameter is also obtained for smooth solutions 
in the context of hyperbolic approximations of high-order PDEs.
The empirical evidence shown in Figure \ref{fig: steady state convergence mbHSGN} 
suggest that such an estimate might also hold for the systems considered in this work,
at least for particularly well-behaved solutions such as \eqref{eq: steady-state EDC}.
However, a rigorous proof of this fact remains an open problem.

For a fixed value of $c^2=50 \si{\meter}^2/\si{\second}^2$ we  perform a convergence test of our numerical scheme
to the steady-state solution of mbHSGN given by \eqref{eq: steady-state EDC}, and the results
are displayed in Table \ref{tab:steady_state_mbHSGN}.
At the highest refinement level, the rate of convergence drops to 1, which we conjecture is due to the discretization
errors interacting with those of the ODE solver in charge of generating the initial condition.

Moreover, we test the ability of our discretization of the HSGN system 
\eqref{eq: HSGN} to capture steady-state solutions of the SGN equations 
\eqref{eq: first order SGN}. For this, we follow Guermond et al. 
\cite{guermond_hyperbolic_2022}, who present the following explicit 
nontrivial steady-state solution for SGN:
\begin{subequations}
  \label{eq: Steady state Guermond}
\begin{gather}
    h(x) = H\left(1+a\,\mathrm{sech}(r x)^2\right),
    \quad b(x) = -\frac{a\, \mathrm{sech}(r x)^2}{2},\\ 
    \quad hu(x) := q=\pm \sqrt{\frac{(1+a)gH^3}{2}},
\end{gather}
\end{subequations}
with $r =H^{-1}\sqrt{3 a/(1+a)}$. The corresponding profiles for the 
auxiliary variables $\sigma, w, p_b, p$ of the first-order reformulation 
\eqref{eq: first order SGN} follow from the SGN constraints by 
straightforward computation; their explicit expressions are reported in 
the supplementary material and used as initial data here.

We fix $c^2=10^5 \si{\meter}^2/\si{\second}^2$ and consider different grid resolutions. The convergence 
results are reported in Table \ref{tab:steady_state_SGN}. As in 
\cite{guermond_hyperbolic_2022}, our nominally second-order scheme attains 
only first-order convergence, which we attribute to the residual 
hyperbolization error at finite $c^2$; confirming this would require a 
semi-implicit discretization able to capture the asymptotic regime 
$c^2\to\infty$ without resolving the fast waves, and is left for future work.

\begin{figure}
    \centering
        \centering
    \includegraphics[width=\linewidth]{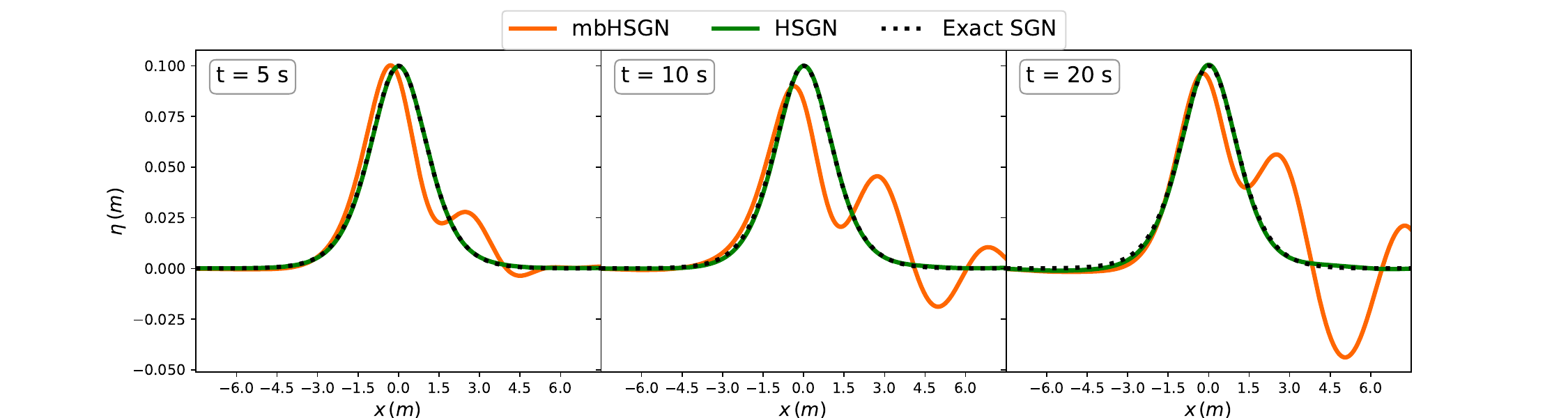}
    \caption{Comparison of the steady-state solutions to the SGN system \eqref{eq: Steady state Guermond}
    obtained with HSGN \eqref{eq: HSGN} and mbHSGN \eqref{eq:mbHSGN}.
    For this test, we set $H=1 \si{\meter}$, $a=0.2 \si{\meter}$, $c^2=1000 \si{\meter}^2/\si{\second}^2$, and $\Delta x=0.005 \si{\meter}$.}
    \label{fig: steady state comparison HSGN vs mbHSGN}
\end{figure}

The explicit solution \eqref{eq: Steady state Guermond} was introduced in \cite{guermond_hyperbolic_2022}
to illustrate the effect of the mild-bottom assumption. Figure \ref{fig: steady state comparison HSGN vs mbHSGN}
compares the two: HSGN captures the SGN steady state with high accuracy, while mbHSGN shows significant discrepancies.
Nevertheless, mbHSGN performs very well in the more general dynamical scenarios of Section \ref{sec:validation}.
\begin{table}[]
    \centering
\begin{tabular}{|c|c|c|c|c|c|c|c|c|}
    \hline
    {N} & 1280 & 2560 & 5120 & 10240 & 20480 & 40960 & 81920 \\
    \hline
    {$L^\infty$ Error $h$} & 0.387 & 0.279 & 0.103 & 0.018 & 0.000947 & 0.000285 & 0.000142 \\
    \hline
    {Rate} & - & 0.47 & 1.43 & 2.52 & 4.25 & 1.73 & 1.00 \\
    \hline
\end{tabular}
    \caption{Errors and rates of convergence for the steady-state solution to mbHSGN \eqref{eq: steady-state EDC} with $c^2 =50\si{\meter}^2/\si{\second}^2$, and $t=1\si{\second}$}
    \label{tab:steady_state_mbHSGN}
\end{table}
\begin{table}[]
    \centering
\begin{tabular}{|c|c|c|c|c|c|c|c|c|}
    \hline
    {N} & 320 & 640 & 1280 & 2560 & 5120 & 10240 & 20480 \\ 
    \hline
    {$L^\infty$ Error $h$} & 0.0661 & 0.0378 & 0.0194 & 0.00984 & 0.00481 & 0.00237 & 0.00118 \\ 
    \hline
    {Rate} & - & 0.81 & 0.96 & 0.98 & 1.03 & 1.02 & 1.01 \\ 
    \hline
\end{tabular}
    \caption{Errors and rates of convergence for the steady-state solution to the SGN
    system \eqref{eq: Steady state Guermond} with $c^2=10^5 \si{\meter}^2/\si{\second}^2$, and $t=1 \si{\second}$}
    \label{tab:steady_state_SGN}
\end{table}

\subsection{Head-on collision of two SGN solitons}
\label{sec: diagonal head-on collision}

\begin{figure}
    \centering
    \includegraphics[width=\linewidth]{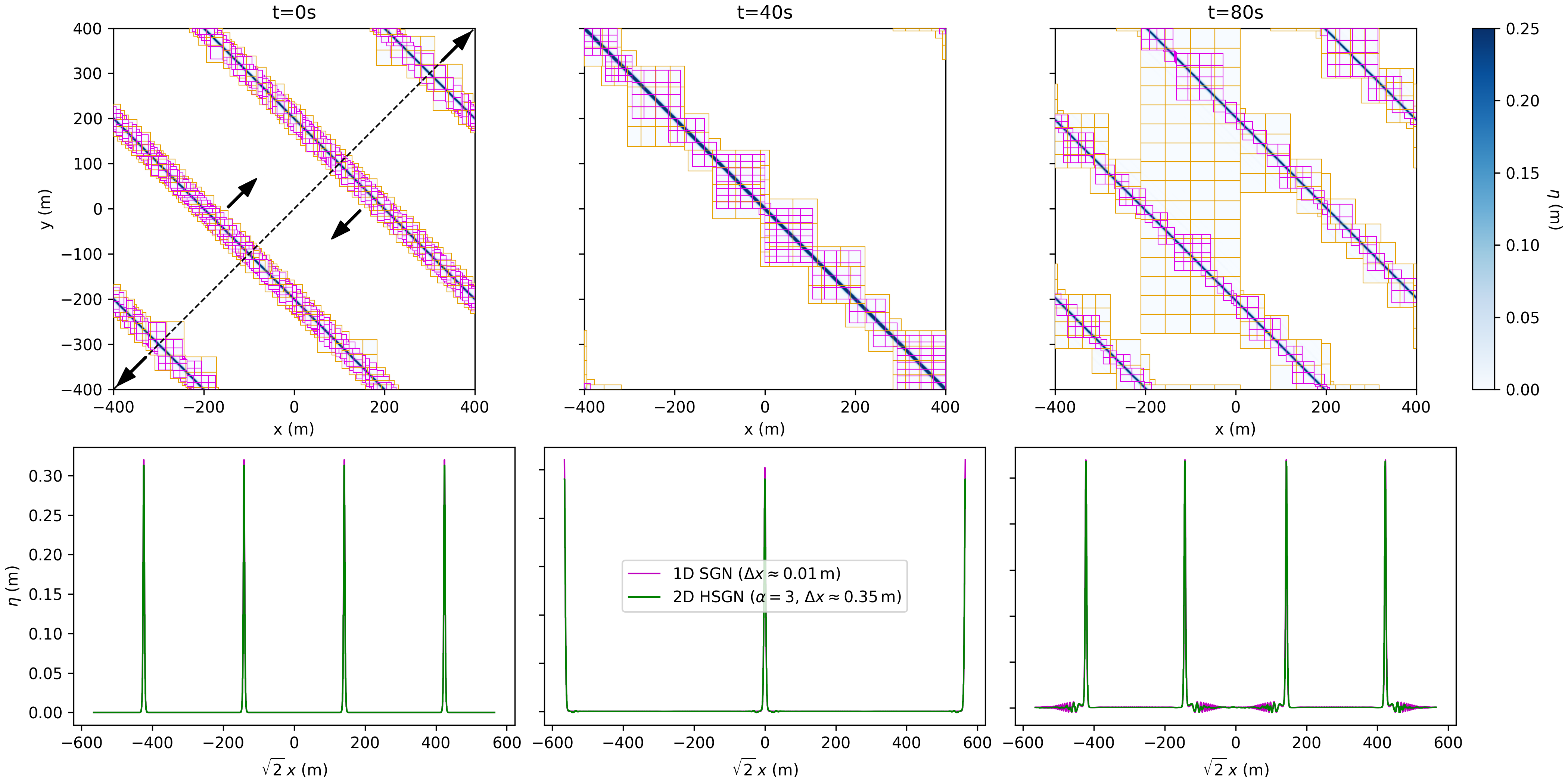}
    \caption{Head-on collision of two SGN solitons (setup described in Section \ref{sec: diagonal head-on collision}). 
    The top row shows the 2D surface elevation at different times with patch boundaries corresponding to levels
    5 and 6 plotted in red and golden respectively. 
    The bottom row shows a slice of the surface elevation along the diagonal $x=y$,
    and compares with a highly resolved 1D reference solution 
    to SGN with periodic boundaries.
    Note the different y-axis ranges in the bottom row.
    \label{fig: 2D SGN soliton collision}}
\end{figure}

With the purpose of validating HSGN as an approximation to SGN,
and to illustrate the performance of our 2D scheme
with AMR,
we consider a quasi-1D scenario unaligned with the grid, in which
the initial condition is given by the superposition
of two SGN solitons of amplitude $a=0.32 \si{\meter}$ over
a flat bottom of depth $H=1 \si{\meter}$
propagating in opposite directions
along the main diagonal of the domain.
Namely, we set 
\begin{align}
{\bfq}(x,y,0) = \begin{cases}
    \tilde{{q}}^+(\xi+d/2), & \text{if } \xi\leq 0 \text{ and } |\xi| \leq d,\\
    \tilde{{q}}^-(\xi+d/2), & \text{if } \xi\leq 0 \text{ and } |\xi| > d,\\
    \tilde{{q}}^-(\xi-d/2), & \text{if } \xi > 0 \text{ and } |\xi| \leq d,\\
    \tilde{{q}}^+(\xi-d/2), & \text{if } \xi > 0 \text{ and } |\xi| > d,
\end{cases}
\end{align}
where $d=\sqrt{2}L/4$ is the initial distance between the soliton peaks,
$\xi=(x+y)/\sqrt{2}$ is the coordinate along the main diagonal,
and $\tilde{q}^+$ and $\tilde{q}^-$ are 1D soliton solutions to SGN 
traveling with positive and negative velocities, respectively (see Supplementary Material for explicit expressions).
The  computational domain is $[-L/2,L/2]^2$, with $L=800 \si{\meter}$ and
periodic boundary conditions.
The coarsest grid has $50 \times 50$ cells, and six levels of refinement
are used in total, with a refinement factor of 2 between each level.
Figure \ref{fig: 2D SGN soliton collision} depicts the results, where strong agreement with a highly 
resolved 1D reference solution to SGN with periodic boundary conditions is observed.

\subsection{Radially symmetric tests}
\label{sec: radial tests}
To further validate our 2D scheme, we consider a radially symmetric scenario in which a Gaussian hump
in the surface elevation is initialized at the origin and allowed to propagate outward.
As in Section \ref{sec: hump flat bottom 1D}, the initial condition is given by the
one-dimensional  function \eqref{eq:initial Gaussian hump} in the radial coordinate with respect to the center of the domain, i.e. $r=\sqrt{x^2+y^2}$,
with $H=100\, \si{\meter}$, $r_0=0\, \si{\meter}$, $a=5\, \si{\meter}$, and $\sigma=100\, \si{\meter}$. 
Well-prepared initial conditions for the additional variables are computed as described in Section \ref{sec:well-prepared}.
Reflecting boundary conditions are imposed at the left and bottom boundaries, while simple zeroth-order extrapolation is used at the right and top boundaries.
Three levels of refinement are used, with an initial grid of $100 \times 100$ cells and two additional levels of refinement
with a refinement factor of 4 and 2, respectively.
The highest level of refinement is only allowed in a square of size $1\, \si{\kilo \meter} \times  1\, \si{\kilo \meter}$ at the origin, while the second level of
refinement is allowed in the region $|x-y| \leq 1\, \si{\kilo \meter}$
(although nearby cells are also refined due to the AMR clustering algorithm \cite{berger_adaptive_1998}).

\begin{figure}
    \centering
    \begin{subfigure}{\textwidth}
      \centering
      \includegraphics[width=\linewidth]{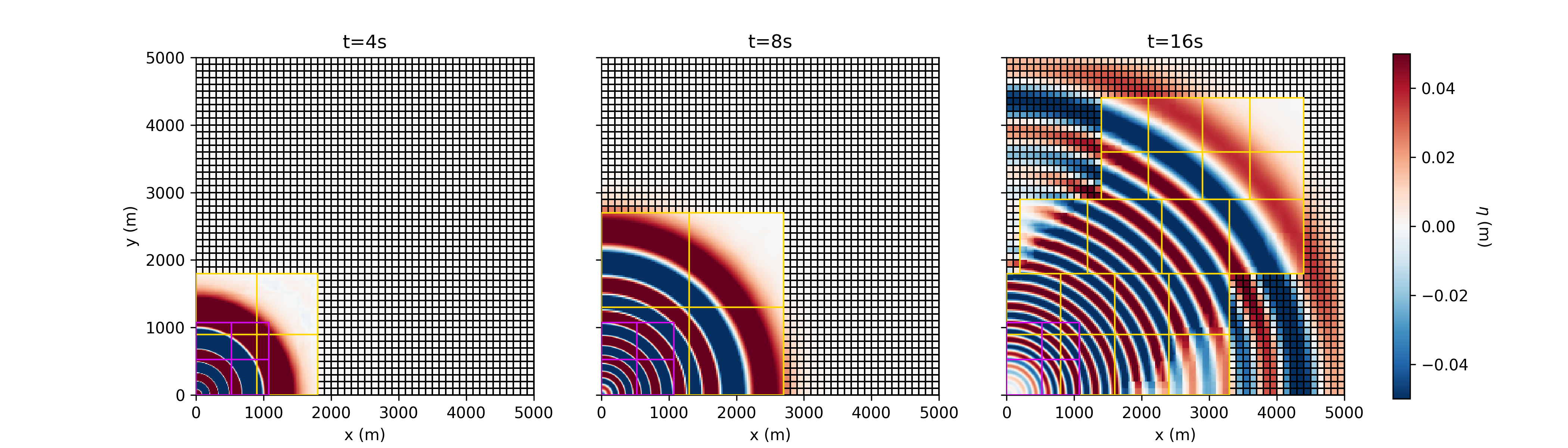}
    \end{subfigure}
    \begin{subfigure}{\textwidth}
      \centering
      \includegraphics[width=0.9\linewidth]{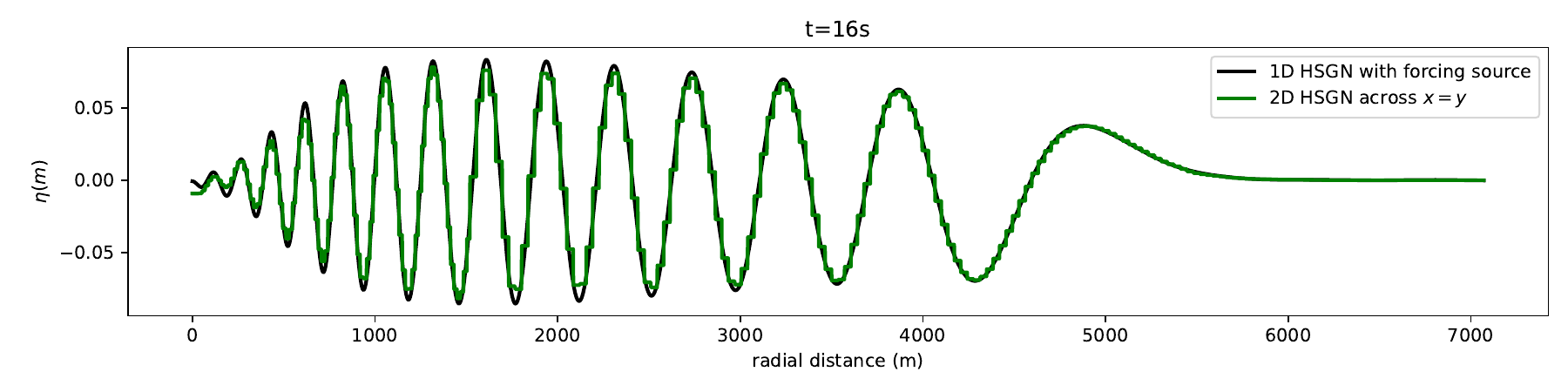}
    \end{subfigure}
    \caption{A Gaussian hump in the surface elevation centered at the origin as described in Section \ref{sec: radial tests}. 
    The top plot shows the 2D surface elevation at different times, 
    while the bottom plot shows a slice of the surface elevation along the main diagonal at $t=16 \si{\second}$,
    compared with a highly-resolved 1D reference solution.
      \label{fig: radial flat}}
\end{figure}

Snapshots of the surface elevation at different times are shown in the top row of Figure \ref{fig: radial flat}, where it is evident that the solution remains radially symmetric
and that the wavetrain is correctly resolved as it propagates outward without artifacts at the interfaces between different levels of refinement.
A comparison of the surface elevation along the main diagonal with a highly resolved 1D reference solution to HSGN is shown in the bottom row of Figure \ref{fig: radial flat},
where excellent agreement is demonstrated.
In order to obtain this 1D reference solution, it suffices to observe that, given the radial symmetry of the problem, the solution to the 2D system \eqref{eq: compact 2D} can be obtained by solving a 1D system 
in the radial coordinate with forcing source terms accounting for the geometrical effects of the polar coordinates
(see supplementary material for details on this reduction).

\section{Validation against experiments and reference solutions}
\label{sec:validation}
Having verified the discretization, we now validate the models against laboratory experiments and a real tsunami event.
The tests are chosen to exercise different aspects of the hybrid approach: dispersive shock formation in a dam-break scenario (Section \ref{sec: undular bores dam break}),
shoaling and run-up over a sloping beach (Section \ref{sec: shoaling run-up}),
wave-obstacle interaction on a mapped grid (Section \ref{sec: wave-obstacle interaction}),
and trans-Pacific propagation in the 2011 Tohoku tsunami (Section \ref{sec:tohoku tsunami}).
The transition to the SWE near the shore, inactive in the previous section, is essential in all four of these tests.

\subsection{Undular bores generated by a dam break}
\label{sec: undular bores dam break}
\begin{figure}
    \centering
    \includegraphics[width=\linewidth]{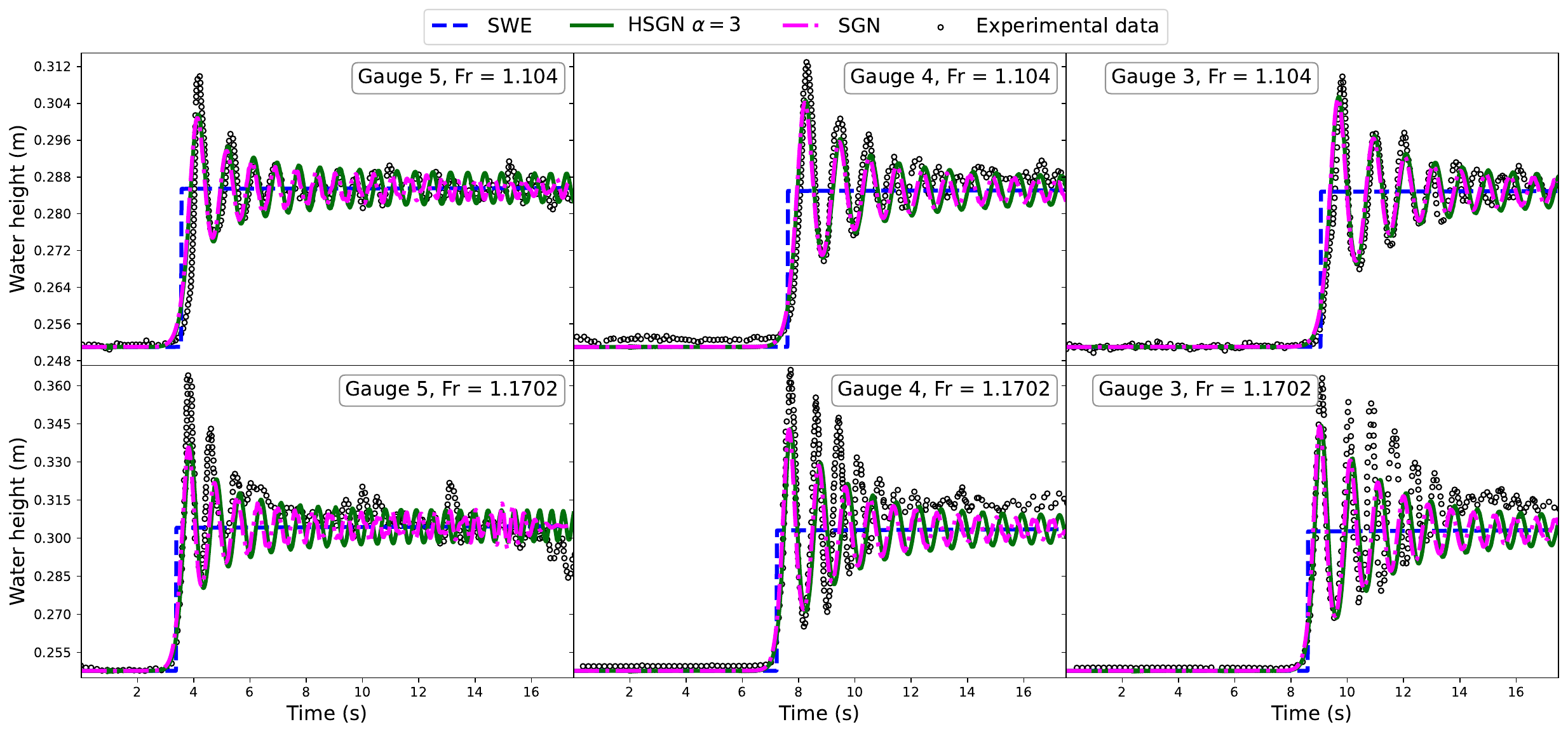}
    \caption{Undular bore generated by a dam break over a flat bottom with different values
    of the Froude number.}
    \label{fig: dam break undular bore}
\end{figure}

To illustrate the ability of our scheme to both handle discontinuous initial data
and capture the generation of dispersive shock waves, we consider
the scenario of a dam break over a flat bottom proposed in \cite{soares2002undular},
for which experimental data is available for the surface elevation.
For the setup and gauge locations, we refer the reader to \cite[Fig. 4]{soares2002undular}.
The initial condition is given by $h(x,0) = H_1+\chi_{x<0} (H_2-H_1)$, $u(x,0)=0$, and $b(x)=0$, where $\chi$ is the indicator function,
and the water depths are chosen such that prescribed values for the upstream Froude number $Fr$ are attained across the undular bore,
i.e. ${H_*}/{H_1} = \frac12 \left(\sqrt{1+8 Fr^2}-1\right)$, where $H_*$ is the depth of the intermediate state between the dam and the undular bore.
Following \cite{parvin_openfoam_2024}, we consider the values $Fr\in\{1.104, 1.1702\}$, which can be approximately obtained by
setting $H_1=0.2477 \si{\meter}$, $H_2=0.3685 \si{\meter}$ and $H_1=0.2510 \si{\meter}$, $H_2=0.3230 \si{\meter}$, respectively.
The surface elevation at different times is shown in Figure \ref{fig: dam break undular bore}, where 
good agreement is found between the numerical solution and the experimental data, especially for the case with $Fr=1.104$.
Moreover, excellent agreement is seen between HSGN and SGN, especially  closer to the leading wave of the undular bore.

\subsection{Matsuyama's wavetank experiment}
\label{sec: shoaling run-up}
In order to simulate the continental shelf, Matsuyama et al. \cite{Matsuyama2007} carried out
an experiment at the $205 \si{\meter}$ long wave flume of the Central Research Institute for Electric Power Industry (CRIEPI) in Japan.
We consider the setup of their experiment in which the beach has slope $1/200$, illustrated in \cite[Fig. 2 (a)]{Matsuyama2007}.
We use the ocean-at-rest solution as initial condition, and
impose non-reflecting boundary conditions for outgoing waves at the left and right boundaries.
To simulate the wave maker, we set the velocity in the left boundary ghost cells to
\begin{align}
u_{BC}(t) =  \begin{cases}\eta_0 \sqrt{g/H} \sin(2\pi t/T), & t<T,\\
0, & t\geq T,
\end{cases}
\end{align}
which gives an incoming wave that is a right-going eigenmode of the linearized SWE.
Here $\eta_0=0.03 \si{\meter}$ is the target amplitude of the surface elevation,
$H=4 \si{\meter}$ is the depth of the ocean at rest at the left boundary, and $T=20 \si{\second}$ is the period of the wave maker.
Transition to the SWE is performed at cells where the water depth is less than $h^*=0.06 \si{\meter}$
to robustly handle the wet-dry front at the right end of the flume.
Our setup follows the 1D GeoClaw demo of the same experiment%
\footnote{\url{https://www.clawpack.org/gallery/gallery/gallery_geoclaw.html\#d-geoclaw-boussinesq-equations}.}.

\begin{figure}
    \centering
    \includegraphics[width=\linewidth]{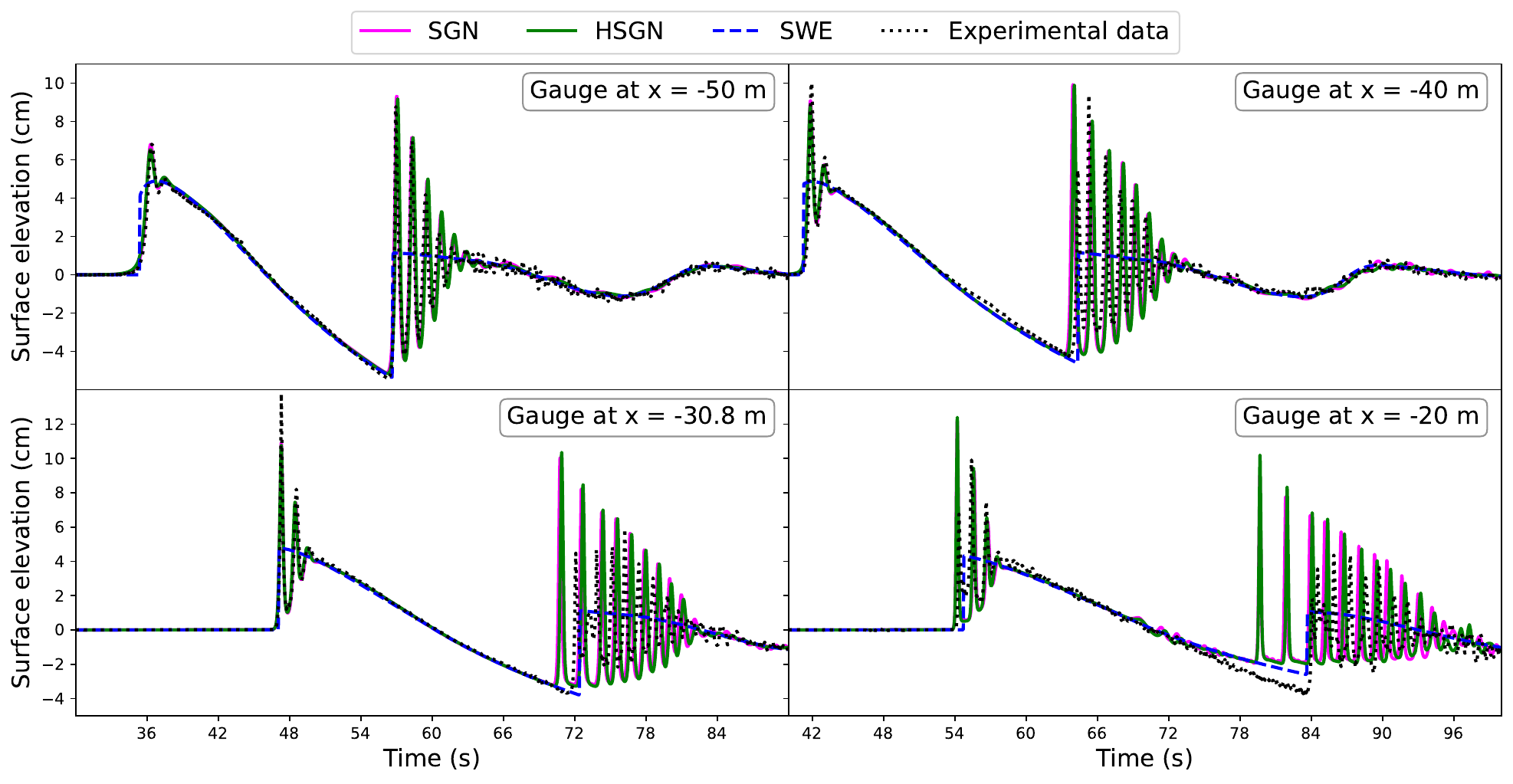}
    \caption{Matsuyama's wavetank experiment: shoaling of a long wave over a sloping beach 
leading to soliton fission. Surface elevation at gauges along the flume, comparing 
HSGN and SWE with the experimental data of \cite{Matsuyama2007}.}
    \label{fig:Matsuyama wavetank}
\end{figure}

Figure \ref{fig:Matsuyama wavetank} shows the surface elevation recorded at different gauges along the flume.
For the sake of clarity, we only show the results obtained with HSGN, since they are
indistinguishable from those obtained with mbHSGN.
Good correspondence is exhibited between the numerical results and the experimental data
at the leftmost gauges, especially in comparison with the results obtained with the SWE.
At the rightmost gauges (the bottom row of Figure \ref{fig:Matsuyama wavetank}), 
discrepancies emerge between the dispersive models and the experimental data
for longer times. 
This reflects the lack of dissipation in the models, since our approach
to trigger wave breaking is based on proximity to the shoreline rather than
on local properties of the flow. The use of more sophisticated wave breaking criteria
could potentially improve the results in this region, as discussed in Section \ref{sec: wave breaking}.
Nonetheless, HSGN tracks SGN closely even in this regime,
indicating that the hyperbolization error is small compared to
the modeling error and that a moderately small relaxation parameter $c^2$ is justified here.

\subsection{Immersed cylindrical obstacle}
\label{sec: wave-obstacle interaction}
\begin{figure}
\centering
\begin{subfigure}{0.49\textwidth}
  \centering
  \includegraphics[width=0.9\linewidth]{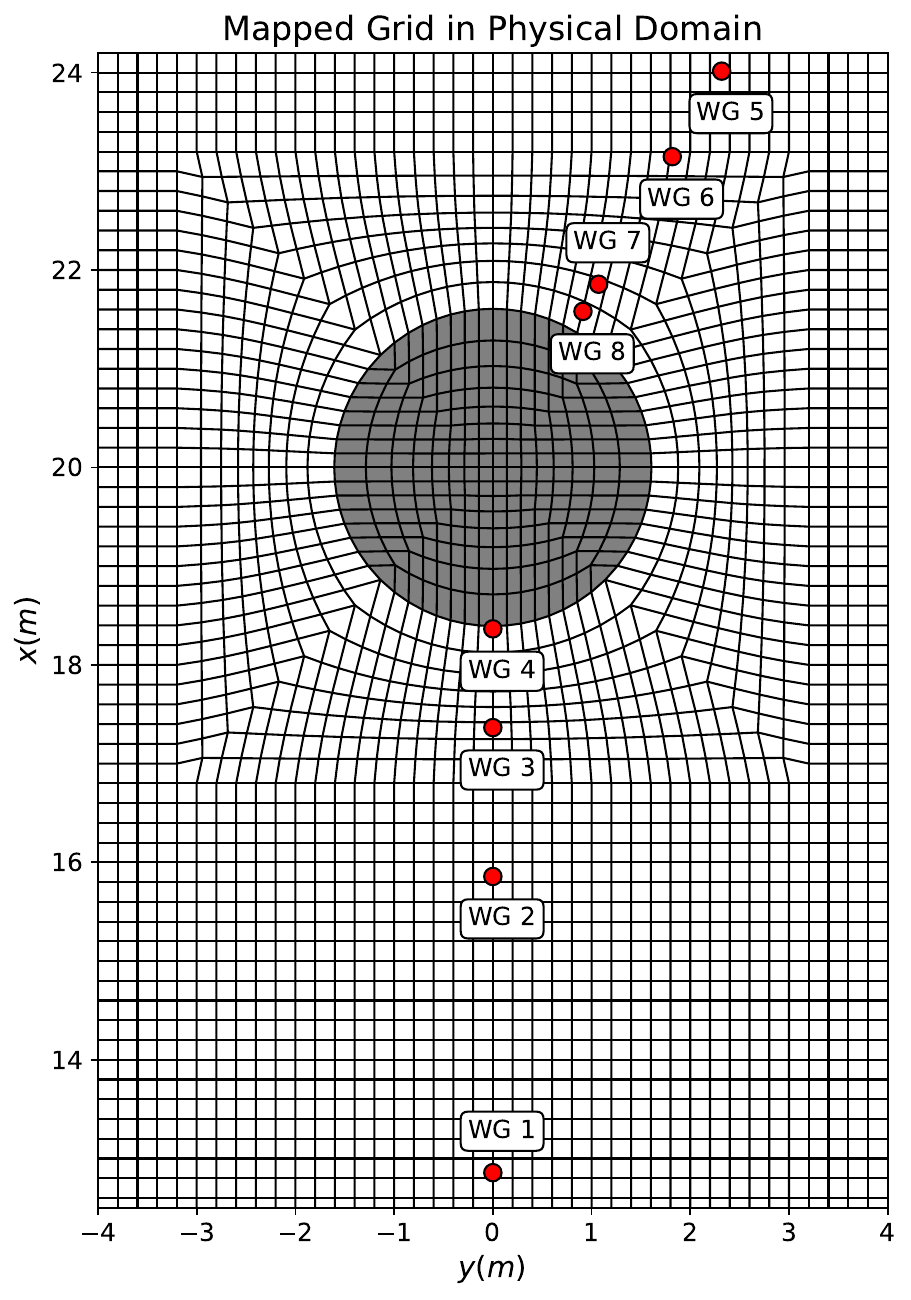}
  \label{fig:cylinder setup}
\end{subfigure}
\hfill
\begin{subfigure}{0.49\textwidth}
  \centering
  \includegraphics[trim={20cm 0cm 20cm 0},clip,width=\linewidth]{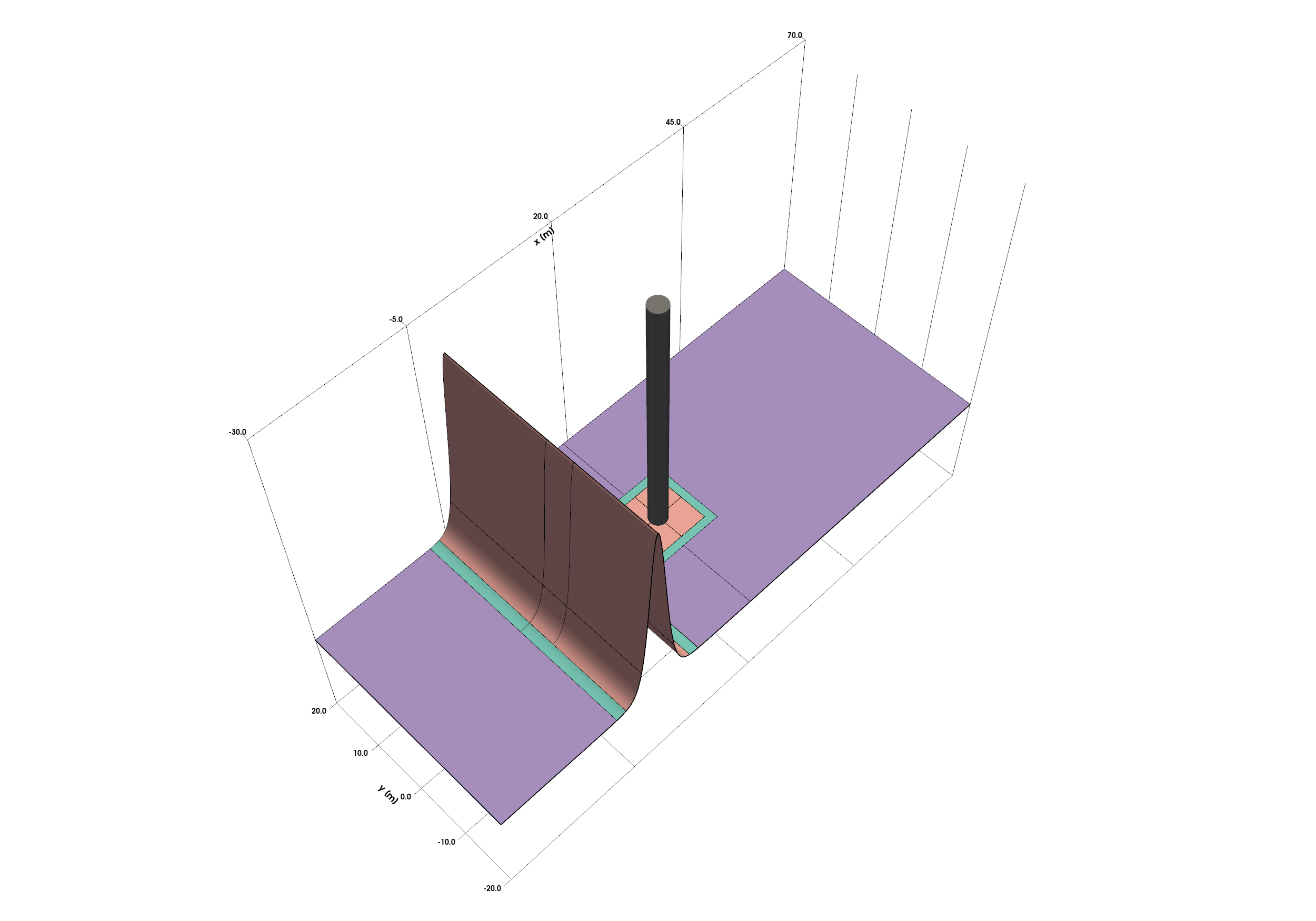}
  \label{fig:cylinder sub1}
\end{subfigure}

\vspace{0.3cm}

\begin{subfigure}{0.49\textwidth}
  \centering
  \includegraphics[trim={20cm 0 20cm 0},clip,width=\linewidth]{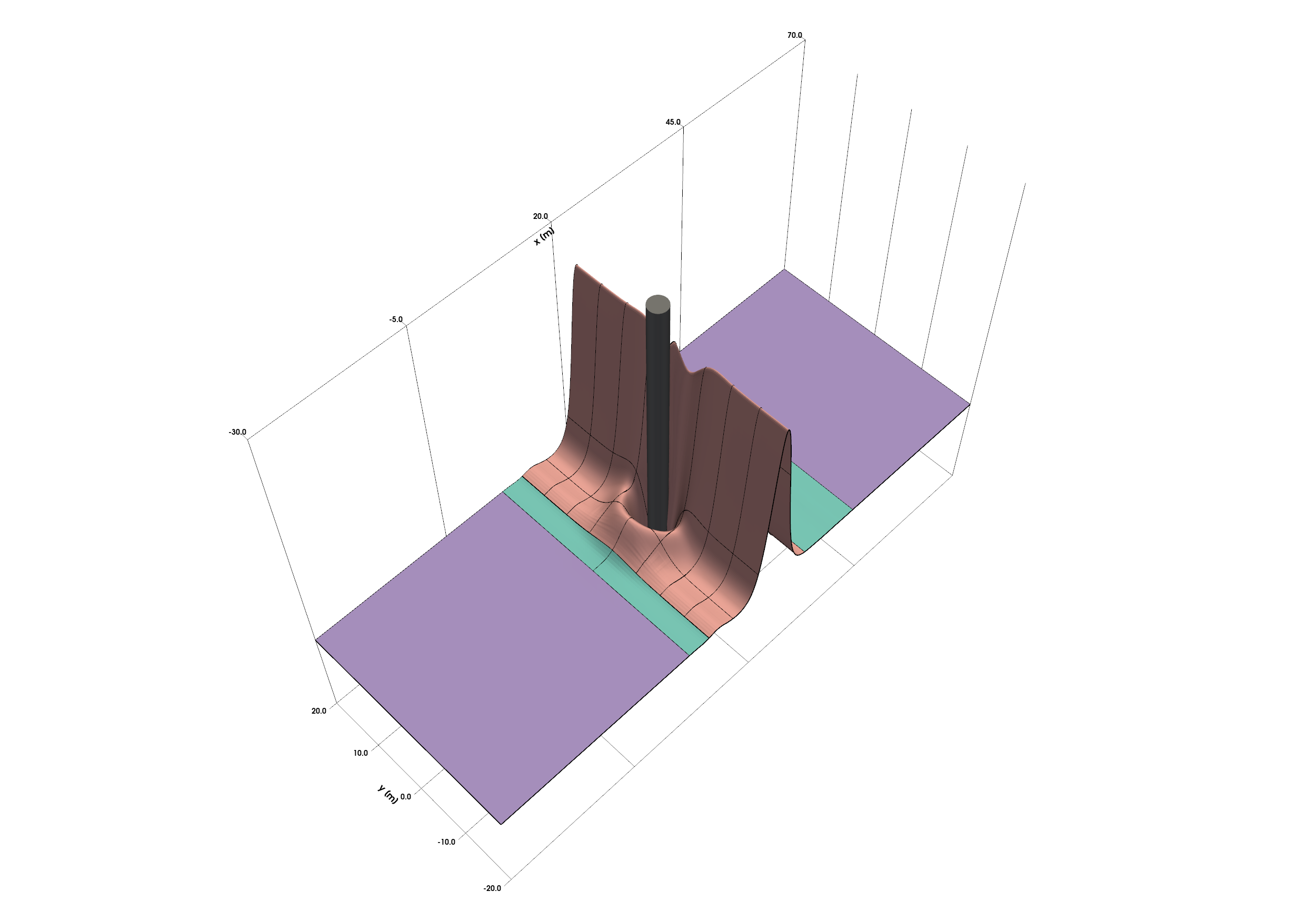}
  \label{fig:cylinder sub2}
\end{subfigure}
\hfill
\begin{subfigure}{0.49\textwidth}
  \centering
  \includegraphics[trim={20cm 0 20cm 0},clip,width=\linewidth]{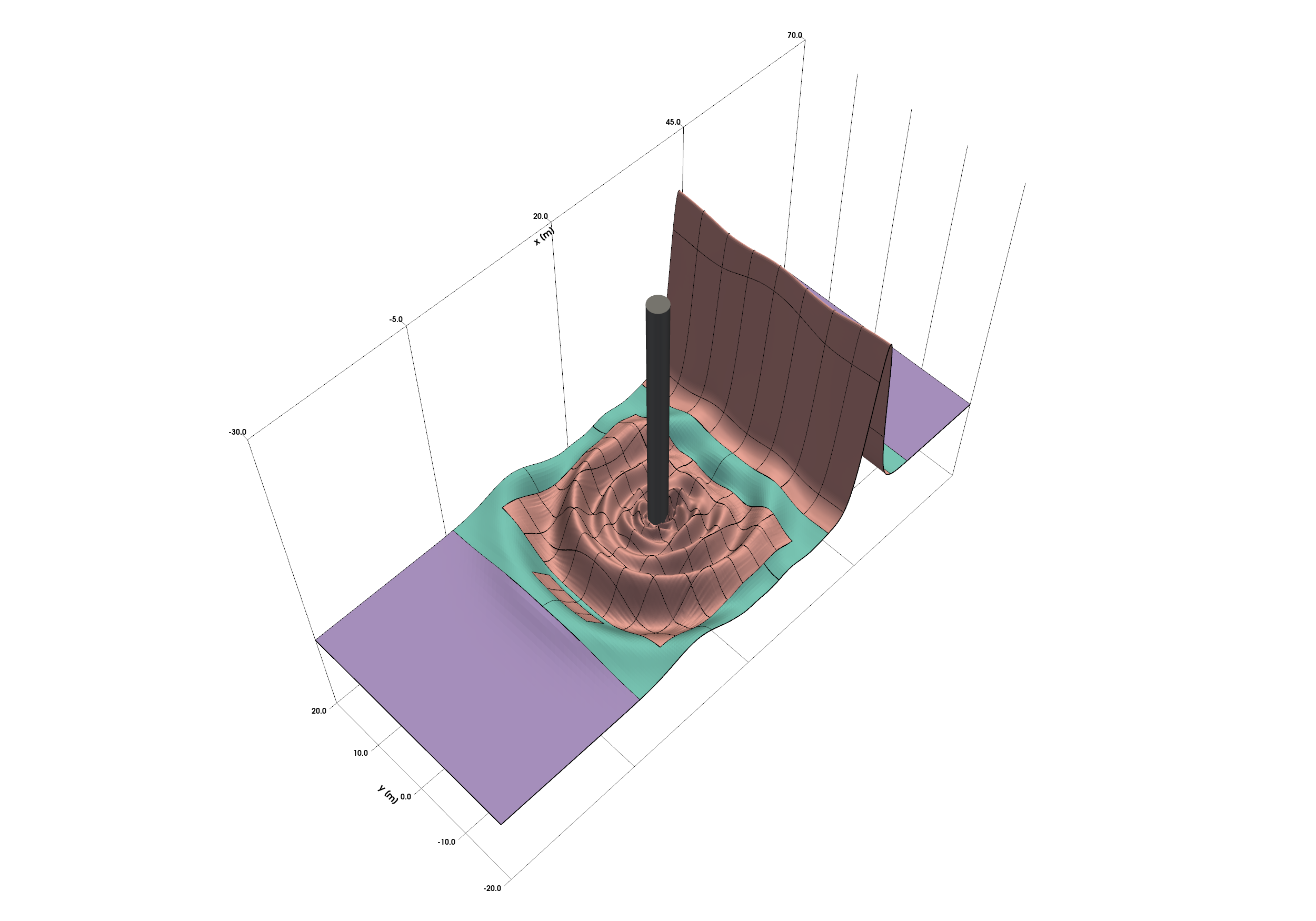}
  \label{fig:cylinder sub3}
\end{subfigure}

\caption{\label{fig:Soliton cylinder 3d surface}
Top left: Single coarse mapped grid (note the rotated axes) around obstacle (in gray), and locations of the gauges (in red).
Top right, bottom left, and bottom right: surface elevation at $t=0 \si{\second}$, $t=6.5 \si{\second}$, and $t=12 \si{\second}$ respectively.
Three levels of refinement are shown, with the coarsest one corresponding to a grid of $110\times 44$ cells 
and increasing the resolution by a factor of 2 and 5 respectively.
Patches corresponding to levels 1, 2, and 3 are shown in lavender, turquoise, and light pink respectively.}
\label{fig:test}
\end{figure}

A significant advantage of working with first-order hyperbolic systems is the ability to
use mapped structured grids to handle complex geometries while retaining the simplicity of a Cartesian grid for, e.g., the 
use of patch-based AMR and transverse Riemann solvers.
To demonstrate this capability, we consider the test problem proposed by Wang et al. \cite{wang_wu_yates_1992},
where a solitary wave propagates over a flat bottom and interacts with a cylindrical obstacle.
For this, we consider (as shown in the top-left panel of Figure \ref{fig:Soliton cylinder 3d surface})
the computational domain $[-30, 70]\, \si{\meter} \times [-40, 40]\, \si{\meter}$, and
a cylindrical obstacle of radius $r=1.5875\, \si{\meter}$, centered at $(x_c,y_c)=(20\, \si{\meter}, 0\,\si{\meter})$.
The boundary of the cylinder is represented exactly in the grid
by use of the mapping proposed in \cite[Sec. 6]{Calhoun_2008}.
The solution is recorded at several gauges, which are also shown in the top left panel of Figure \ref{fig:Soliton cylinder 3d surface}.

The bathymetry is given by $b(x,y)=-1~\si{\meter}$, and the initial condition is
given by a planar (extended along the $y$-axis)  solitary wave solution to SGN of amplitude $a=0.4\, \si{\meter}$
centered at $(x_0,y_0)=(0\, \si{\meter}, 0\,\si{\meter})$, and propagating in the positive $x$-direction.
We consider three levels of refinement with an initial coarse grid of $110 \times 44$ cells, and a refinement 
factor of 4 and 5 for the second and third levels of refinement, respectively, i.e., an effective 
resolution of $2200 \times 880$ is attained in the finest level of refinement.
Cells are flagged for refinement if the difference in surface elevation between neighboring cells is larger than 
$0.05\, \si{\meter}$. Moreover, we force a square of size $3.2 \, \si{\meter} \times 3.2 \, \si{\meter}$ centered at the obstacle to be refined at the finest level at all times
to ensure that the interaction of the wave with the obstacle is well resolved.
At the interior boundary cells corresponding to the obstacle, we prescribe ghost cell values such that the normal
velocity at the boundary is zero, i.e. $\bfu_{G} = [-\bfu \cdot \hat{\bfn},0]^T$, where $\hat{\bfn}$ is the normal vector to the boundary pointing outward from the fluid domain. 

\begin{figure}
    \centering
    \includegraphics[width=\linewidth]{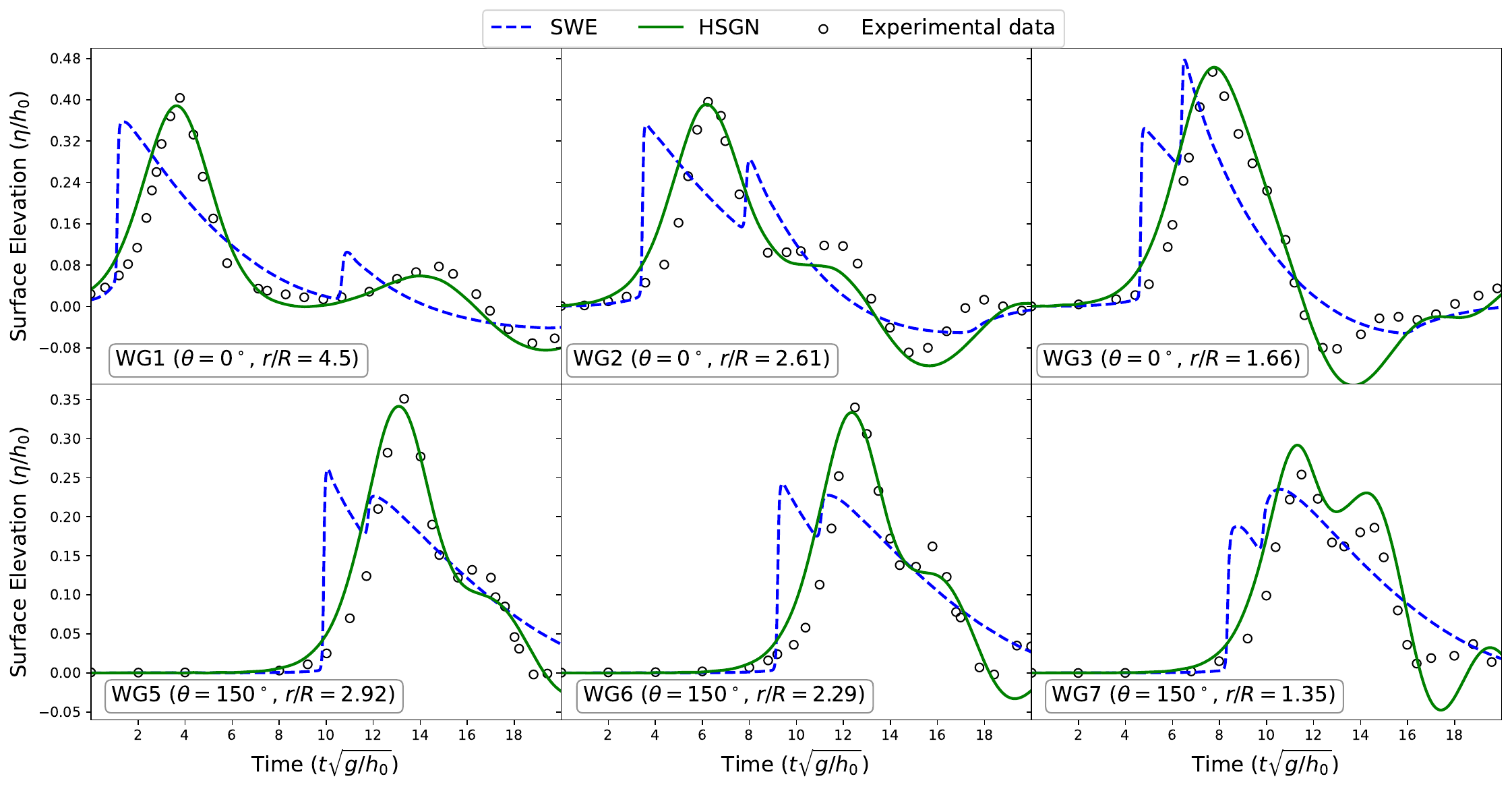}
    \caption{Gauge comparison for the soliton interacting with a cylindrical obstacle test case.
    \label{fig: soliton cylinder gauge comparison}}
\end{figure}

Snapshots of the surface elevation are shown in Figure \ref{fig:Soliton cylinder 3d surface},
illustrating the interaction of the wave with the obstacle.
A comparison with experimental data is shown in Figure \ref{fig: soliton cylinder gauge comparison}.
We see that the HSGN solution matches the data relatively well, while the SWE solution exhibits unphysical
wave breaking and fails to track the data.

\subsection{Tohoku tsunami}
\label{sec:tohoku tsunami}
As a real-world validation, we simulate the Tohoku tsunami of 11 March 2011, 
generated by a magnitude 9.0 megathrust earthquake off the Pacific coast 
of Japan. We adopt the setup of \cite{berger_implicit_2023}: the seafloor 
deformation is prescribed by the  source 
models of Fujii et al. \cite{fujii2011tsunami} and Shao et al. \cite{shao2011focal} (called Fujii and UCSB models respectively,
following the naming convention of \cite{berger_implicit_2023}),
the topography, computational domain, 
DART buoy locations, AMR levels (six in total, from $2^\circ \approx 220\,\si{\kilo\meter}$ 
on level 1 down to $1/3$ arcsecond $\approx 10\,\si{\meter}$ in level 6,
with refinement ratios $5,6,4,6,30$), and refinement criteria 
are all as described there. The only methodological differences are the 
governing model (mbHSGN/HSGN in place of SGN) and the corresponding 
discretization described in Section~\ref{sec:methods}. For the model 
transition we choose the water depth threshold 
$h^*=20\,\si{\meter}$, and a Courant number of 0.75 is used for all the simulations.
The initial condition is given by the ocean-at-rest solution,
and the tsunami is subsequently generated through sea-floor motion.
Zeroth-order extrapolation is used to approximate non-reflecting conditions at the boundaries of the computational domain.
Additionally, a simple scalar relaxation-based
sponge layer \cite{munoz2026unified} of one degree
of width is used to minimize reflections
at the bottom and right boundaries.

\begin{figure}
    \centering
    \includegraphics[trim={2cm 2cm 1cm 0},clip,width=\linewidth]{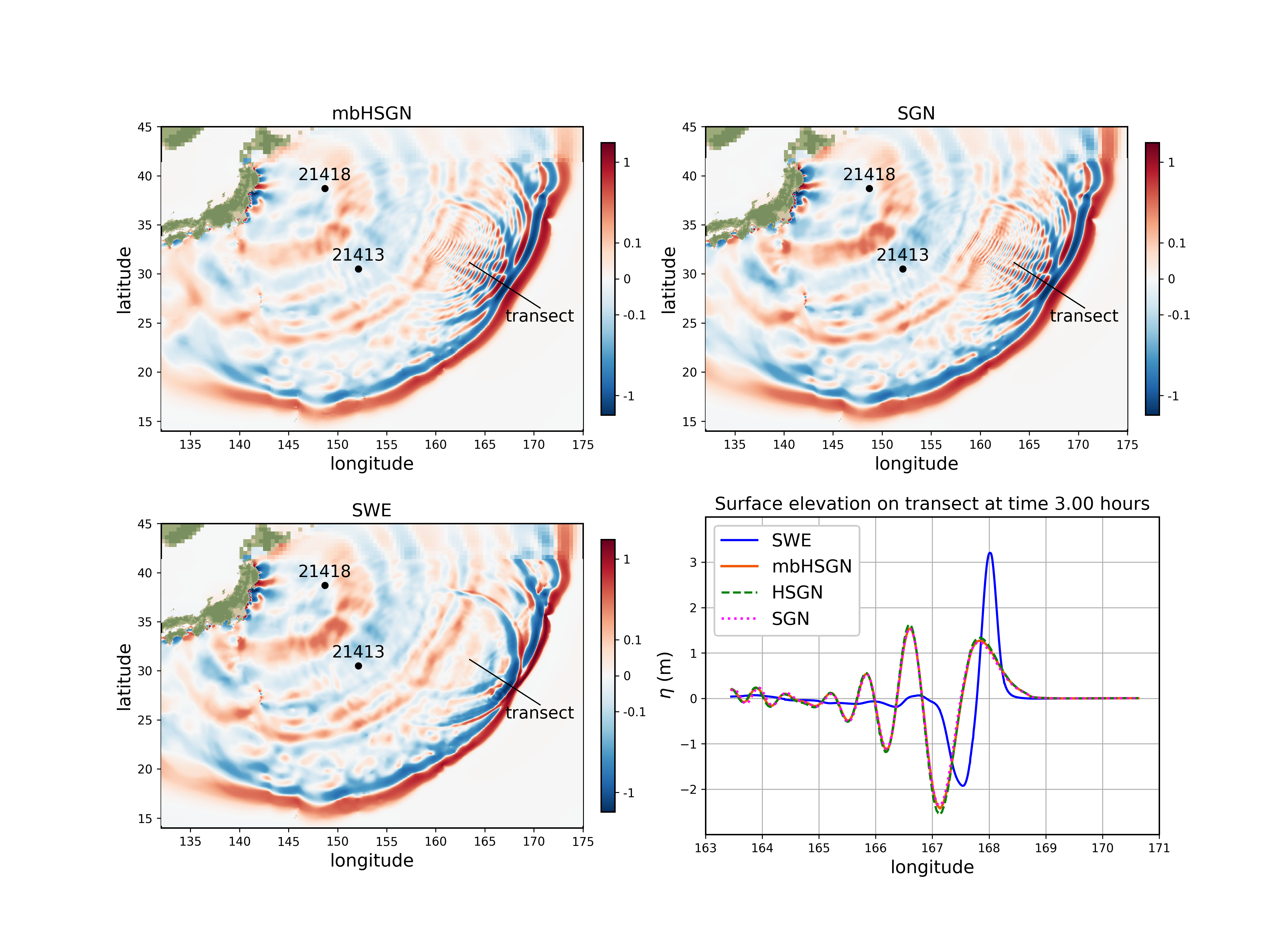}
    \caption{Surface elevation of the Tohoku tsunami close to the shore of Japan after 3 hours of propagation using
    the UCSB source model}
    \label{fig: Tohoku tsunami}
\end{figure}

\begin{figure}
    \centering
    \begin{subfigure}[b]{0.495\textwidth}
    \centering
    \includegraphics[width=\linewidth]{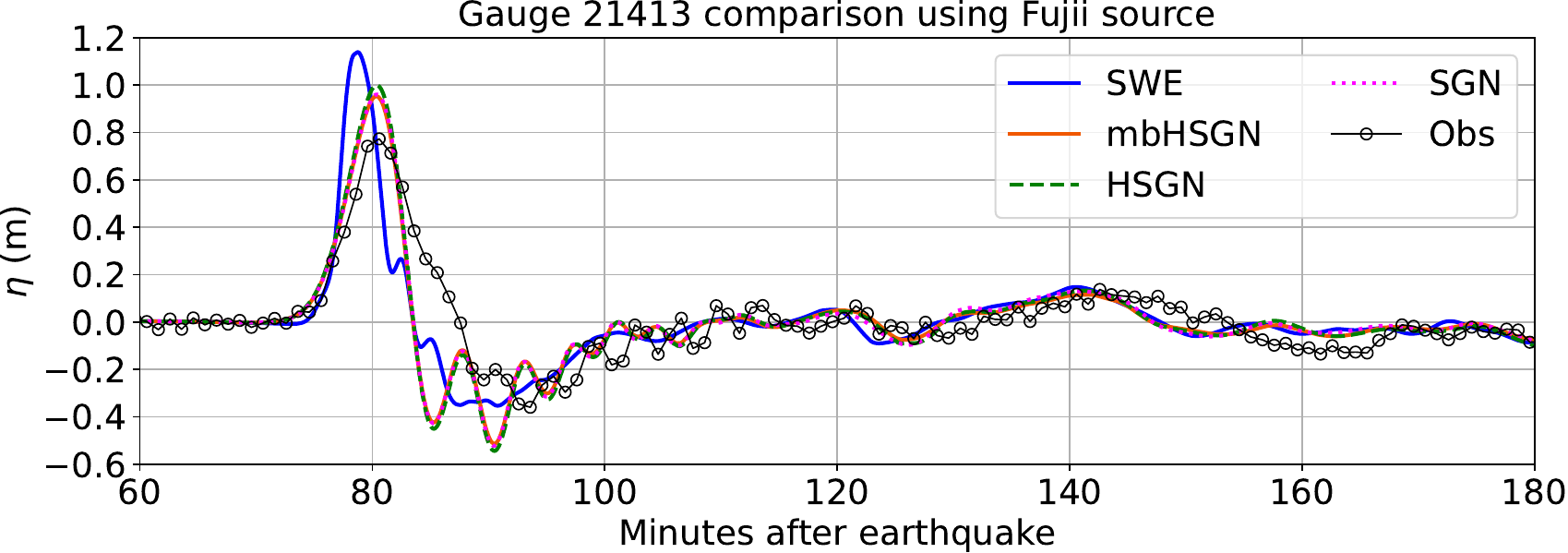}
    \label{fig:tohoku DART 3}
    \end{subfigure}
    \hfill
    \begin{subfigure}[b]{0.495\textwidth}
    \centering
    \includegraphics[width=\linewidth]{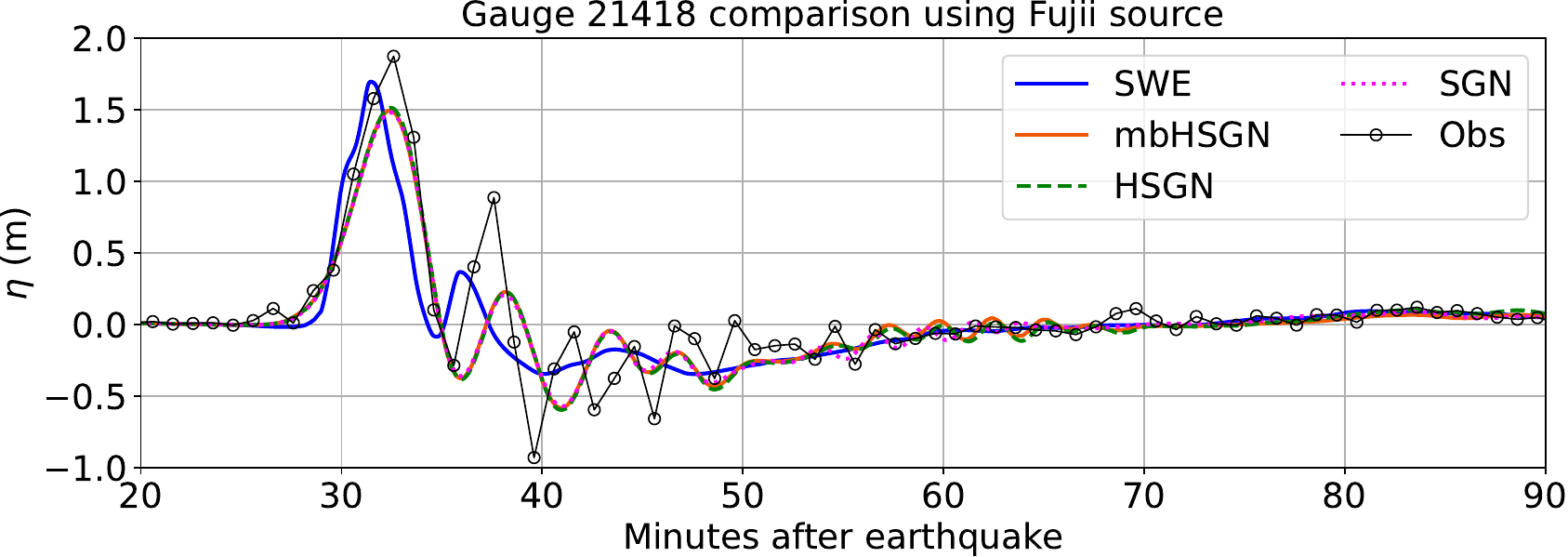}
    \label{fig:tohoku DART 8}
    \end{subfigure}%
    \caption{Comparison with DART buoy data for the Tohoku tsunami. Left: DART 21413. Right: DART 21418.
    \label{fig:tohoku DART comparison}}
\end{figure}

Figure \ref{fig: Tohoku tsunami} shows snapshots of the surface elevation 
during propagation over the Pacific after 3 hours, along with a comparison 
across the transect from (163.454E, 31.1458N) to (170.631E, 26.5282N),
proposed in \cite{POPINET2015336}.
Comparison with the DART 21413 and 
21418 records is shown in Figure \ref{fig:tohoku DART comparison}. 
The results obtained with mbHSGN
and HSGN are very close to those obtained with SGN,
confirming that, as in the previous benchmarks,
the hyperbolization error is negligible compared to the modeling error in this regime and time scale.

While the discretization described in Section \ref{sec:methods} is 
robust enough to handle a wide range of realistic bottom topographies,
extremely large jumps in the bathymetry might still lead to spurious oscillations
in some isolated cells after the still water gets perturbed by the flow
(nearby the Hawaiian Ridge in this test).
We have found HSGN to be more susceptible to this issue than mbHSGN, and
that locally forcing a transition to the SWE suffices to mitigate it.

\section{Computational efficiency}
\label{sec: cost}
As discussed in the introduction, a primary motivation for the use of
dispersive hyperbolic models is the prospect of improved computational
efficiency.  The present work provides an ideal opportunity to evaluate
whether this prospect can be achieved in practice, since we can compare
our solvers with the existing SGN solver that is implemented in the same
computational framework and languages, on problems defined in precisely
the same way.  To that end, we revisit one idealized example and one
real-world example: the radial hump of Section \ref{sec: radial tests}
and the Tohoku tsunami of Section \ref{sec:tohoku tsunami}.

\begin{figure}
\centering
\begin{minipage}{.49\textwidth}
\centering
  \includegraphics[width=\linewidth]{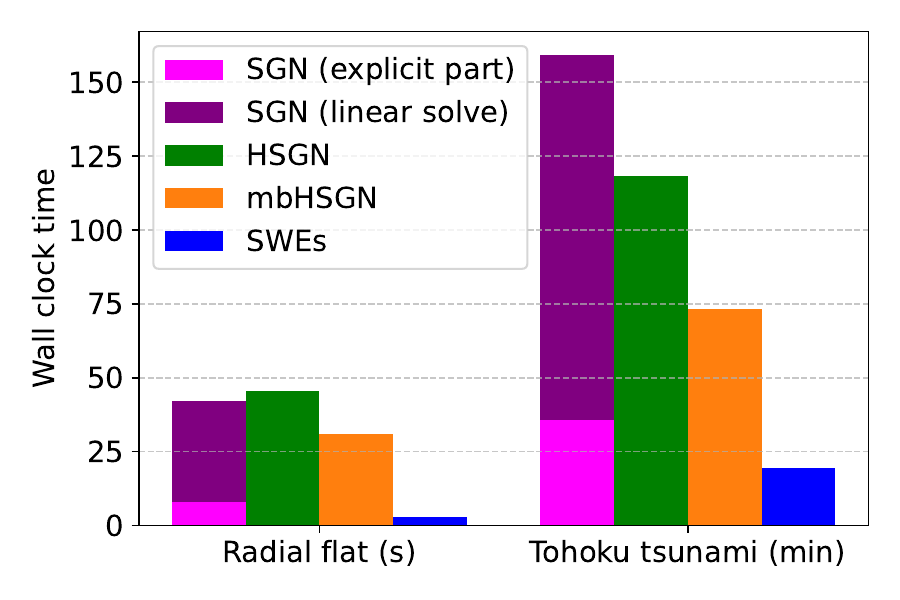}
  \captionof{figure}{Wall-clock time of the different examples.}
  \label{fig: Wall-clock time several examples}
\end{minipage}%
\hfill
\begin{minipage}{.49\textwidth}
  \centering
  \includegraphics[width=\linewidth]{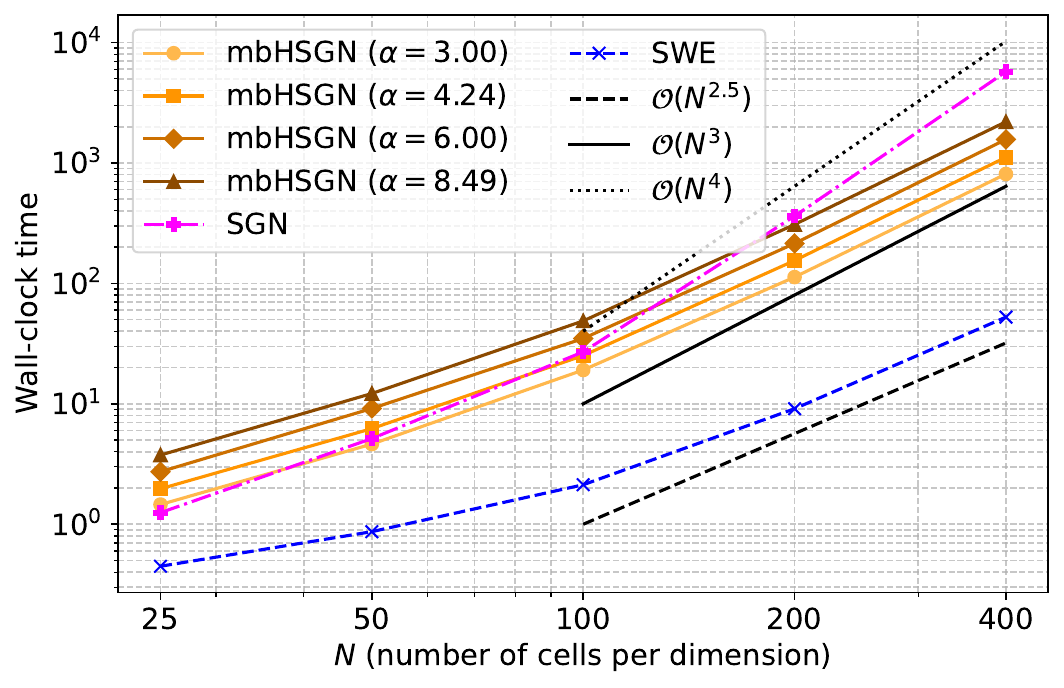}
  \captionof{figure}{Wall-clock times of mbHSGN, SGN, and SWE for different initial refinements and 
  linearly increasing relaxation parameters $(c^2 = \alpha^2 g H)$}
  \label{fig: mbHSGN SWE ratio heatmap}
\end{minipage}
\end{figure}

We measure the total wall-clock time running with 8 OpenMP threads (physical cores)
on an AMD Ryzen Threadripper 7980X CPU.
For SGN, the elliptic problem at each step is solved using GMRES with HYPRE's BoomerAMG as a preconditioner
on an optimized PETSc build.
The results are shown in Figure \ref{fig: Wall-clock time several examples}. 
The first-order dispersive hyperbolic solvers provide a substantial advantage over the
high-order dispersive SGN solver, particularly for the realistic tsunami scenario, in
which the cost of mbHSGN is roughly half of the cost of SGN.   Although the hyperbolic
models require more time steps (as they include faster waves related to the relaxation parameter),
this cost is still smaller than that of the large linear algebraic solves required for SGN.

The non-dispersive SWE solver is dramatically faster than any of the dispersive solvers,
for two reasons.  The first is that it requires no algebraic solves and has no artificial fast waves.
The second reason is more subtle but very significant: in the SWE solution, a much smaller
region of the grid is refined, especially in the radial hump scenario.  That is because
the dispersive models generate trailing oscillations behind the leading wave, while the SWEs
leave a nearly undisturbed surface behind.
Therefore, when using AMR, there is an associated increase in the cost of
mbHSGN (as well as HSGN and SGN) with respect to the SWEs that is unavoidable
due to the physics of the problem modeled by these systems.
In practical applications, this cost can be significantly reduced by deciding which regions of the domain require refinement,
for instance by only tracking waves approaching the shore rather than aiming to resolve every feature of the solution.

For the case of a radial hump over a flat bottom, we further explore how
the computational cost (wall-clock time) of mbHSGN  varies
as we refine the grid and increase the relaxation parameter $c^2(\alpha)$ (and thus the wave speeds of the hyperbolic model).
The results are shown in Figure \ref{fig: mbHSGN SWE ratio heatmap},
where we can estimate the relative cost of mbHSGN to grow as
\begin{align}
{T_\text{mbHSGN}}(\alpha, \Delta x) \approx \mathcal{O}(\alpha N^{2.8}),
\end{align}
while for SGN we have
\begin{align}
{T_\text{SGN}}(\Delta x) \approx \mathcal{O}(N^{4}),
\end{align}
where $N$ is the number of cells per dimension.
These estimates are problem-dependent and will vary with the discretization and refinement strategy,
but they give a useful guide for choosing the relaxation parameter to balance cost and accuracy
(with respect to the SWEs and SGN) of mbHSGN 
and HSGN for a given grid refinement.
Combined with the quadratic growth in $\alpha$ (linear in $c^2$) of the hyperbolization error 
for nontrivial steady solutions (Figure \ref{fig: steady state convergence mbHSGN}),
the linear cost scaling in $\alpha$  suggests an optimal value that minimizes
the error/cost ratio relative to SGN, a question left for future work.

\begin{figure}
\centering
\begin{minipage}{.49\textwidth}
\centering
  \includegraphics[width=\linewidth]{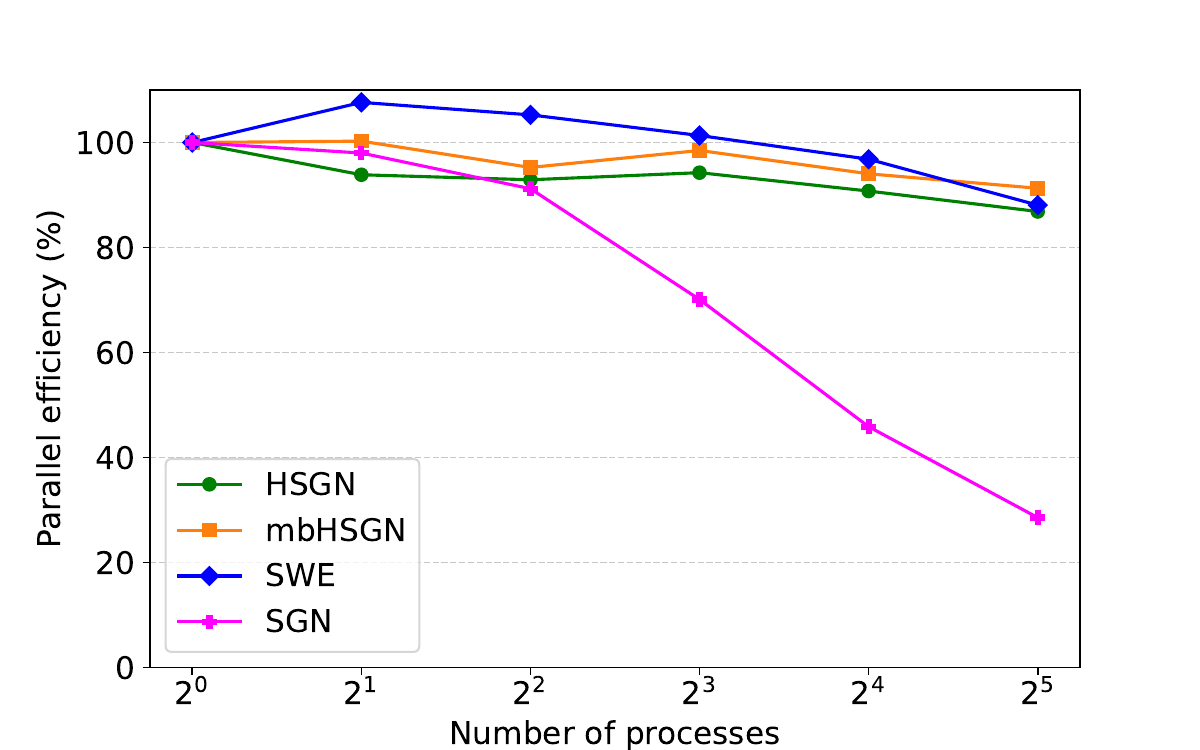}
  \captionof{figure}{Weak scaling comparison for the radial flat test case using a fixed grid.}
  \label{fig: Weak scaling comparison fixed grid}
\end{minipage}%
\hfill
\begin{minipage}{.49\textwidth}
  \centering
  \includegraphics[width=\linewidth]{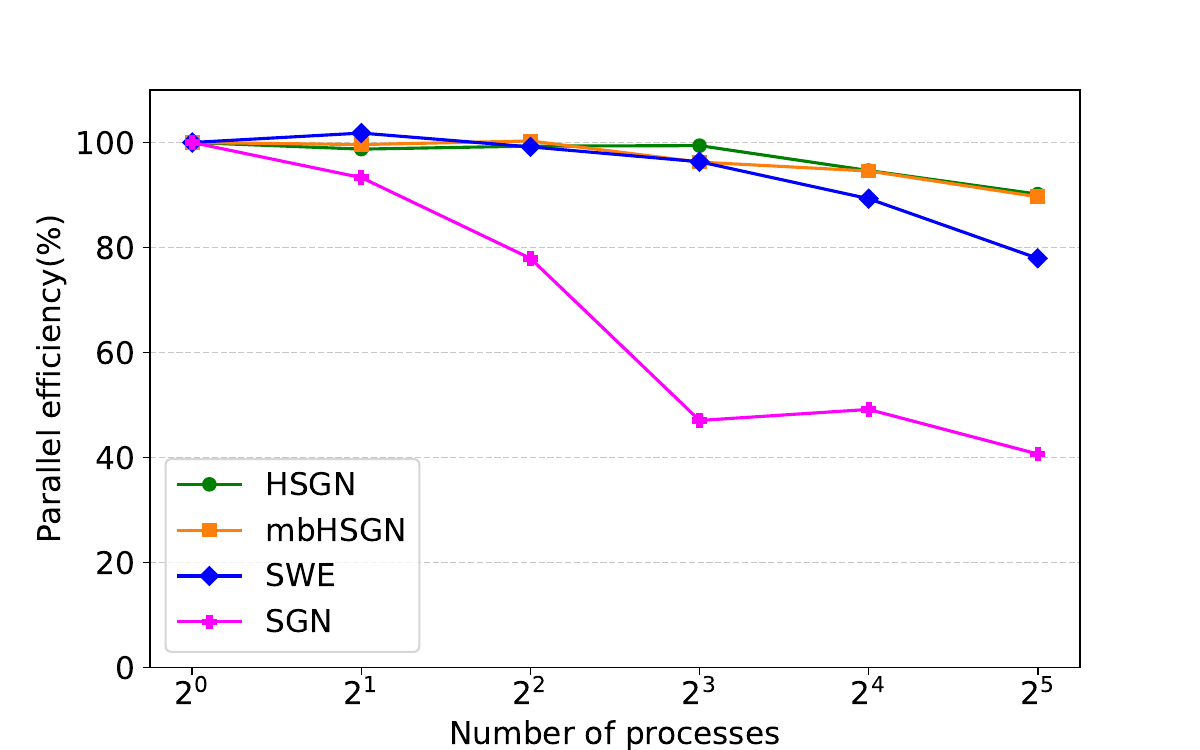}
  \captionof{figure}{Strong scaling comparison for the radial flat test case using a fixed grid.}
  \label{fig: Strong scaling comparison fixed grid}
\end{minipage}
\end{figure}

A preliminary analysis of the weak and strong scaling of the different models is shown in Figures
\ref{fig: Weak scaling comparison fixed grid} and \ref{fig: Strong scaling comparison fixed grid}. 
In both cases we choose $\alpha=3$ and use the radial flat test described in 
Section \ref{sec: radial tests} with a fixed cell size of $\Delta x = \Delta y=12.5 \si{\meter}$.
For the weak scaling comparison, we increase the length of the domain and the final time of the simulation as we
increase the number of processes
so as to maintain a constant workload per process. For a two-dimensional hyperbolic problem on a domain of size $L\times L$, the work per step
grows as $\mathcal{O}\!\left((L/\Delta x)^2\right)$ and the total number of steps grows as 
$\mathcal{O}(T_{\text{final}}/\Delta t)=\mathcal{O}(T_{\text{final}}/\Delta x)$, giving total work
$\mathcal{O}(L^2 T_{\text{final}}/\Delta x^3) \approx \mathcal{O}(L^3/\Delta x^3)$ for a fixed ratio $T_{\text{final}}/L$.
Thus, for fixed $\Delta x$, the work per process is approximately constant provided $L \propto N_{\text{proc}}^{1/3}$.
For the strong scaling comparison, the domain size, mesh width, and final time are held fixed.

For both weak and strong scaling, the first-order hyperbolic solvers (both SWE and the dispersive models)
scale with high efficiency almost up to 32 processes.
Since these algorithms are explicit, the only obstacle to scaling is increased communication
of ghost cell values and wave speeds (for the CFL-based step size controller).
Remarkably, they even scale better than the SWE solver in some cases,
which we attribute to the higher arithmetic intensity of the
Riemann solver and the fully local algebraic source-term integration, both of which are favorable in a multi-threaded setting.
In contrast, the SGN solver rapidly loses efficiency due to the more difficult task of scaling
the algebraic solve.

In summary, we find that the first-order dispersive hyperbolic solvers provide a computational advantage
over the SGN solver, and that advantage is more pronounced for realistic problems, finer grids, and larger
parallel computing systems.

\section{Conclusion and future work}
We have presented a fully explicit, hybrid hyperbolic-dispersive solver for tsunami modeling, implemented within the GeoClaw framework.
The method combines the shallow water equations with either of two hyperbolic approximations of the Serre-Green-Naghdi equations, the full-bottom HSGN or its mild-bottom
variant mbHSGN, discretized using a wave-propagation formulation of Godunov-type methods on logically rectangular grids with
patch-based adaptive mesh refinement.
Dispersive effects are retained in the deep ocean and a transition to the shallow water equations near the shore handles wave breaking and run-up,
with the relaxation parameter varying in space to avoid prohibitive stiffness in shallow water.

The immediate advantage of this approach is that the entire system of equations is first-order
hyperbolic; thus no algebraic solver is required, and enhancements like parallelization and
adaptive mesh refinement can be applied in a relatively simple and standard way.

Since the hyperbolic system involves an additional approximation (with the associated hyperbolization error), 
the first concern regarding this approach is whether it can provide comparable or better accuracy
to what is obtained with a high-order dispersive solver. 
Since the hyperbolic system introduces additional fast waves not present in the shallow water equations,
the second concern is whether it can be implemented at a reasonable computational cost.  

We have seen through both analysis and extensive examples that the hyperbolization error can be
made small (smaller than the modeling errors) with even a moderate value of the relaxation parameter,
allowing the use of a fully explicit time discretization and ensuring that the CFL constraint
due to the additional waves is not too restrictive.  As seen in Section \ref{sec: cost}, the
resulting method is in most cases even more efficient than solving the original dispersive (SGN) model,
and provides as much as a 2x speedup on a real-world tsunami example.

The question of whether dispersive first-order hyperbolic discretizations can
outperform existing high-order dispersive discretizations is of great importance
not only for water wave applications but for a large variety of physical problems
modeled by high-order PDEs.  The above-mentioned concerns regarding accuracy and
cost have so far prevented the adoption of these methods in applications.
In the present work, we have dealt with highly realistic scenarios
in complicated multidimensional geometries and applied advanced computational tools
including adaptive mesh refinement and shared-memory parallelism.
Furthermore, we compared the first-order approach directly with a solver
for the original high-order model, implemented in the same languauge and framework.
We believe this is the most convincing example yet in showing that
first-order hyperbolic relaxation can sometimes provide a more efficient approach for practical applications.

The mild-bottom model, despite its inability to capture the SGN steady state, 
performed very well in dynamical scenarios at roughly half the cost of SGN, making it an attractive practical compromise.
Although the current implementation is parallelized only on shared-memory architectures,
a preliminary scaling study showed that the dispersive hyperbolic models scale at least as well as the SWE,
which places the extension to a distributed-memory parallel implementation, in e.g. the ForestClaw version of GeoClaw,
as a natural next step.
Extending the framework to hyperbolic approximations with improved dispersion in deep water is another 
natural direction for future work.
Further validation of the solver, in the presence of complex geometries
and large-scale realistic and hypothetical tsunami scenarios, is the subject of current work and will be reported in a forthcoming paper.

\section*{Acknowledgments}
We would like to thank Marsha Berger, Randall LeVeque, and Kyle Mandli for their continuous support and feedback 
during the development of this work. 
We also thank Cipriano Escalante, Donna Calhoun, and Dimitrios Mitsotakis for
helpful discussions and suggestions regarding the implementation and validation of the solvers.

\bibliographystyle{plain}
\bibliography{references}
\newpage
\appendix
\section{Well-prepared initial data for SGN steady state}
Well-prepared initial data for the additional variables of mbHSGN and HSGN can be obtained explicitly for
the steady-state solution of SGN introduced in \cite{guermond_hyperbolic_2022}:
\begin{subequations}
\label{eq: Steady state Guermond extra vars}
\begin{align}
    \sigma(x) &= -h(x) u'(x) = -\frac{4\, a\, q\, r \tanh (r x)}{2 a+\cosh (2 r x)+1},\\
     w(x)&=\frac{\sigma(x)}{2}+u(x)b'(x)
    =-\frac{2 a (H-1) q r \tanh (r x)}{H (2 a+\cosh (2 r x)+1)},\\
    p_b(x)&=q w'(x)=\frac{a (H-1) q^2 r^2 \text{sech}^2(r x) (-4 a-2 \cosh (2 r x)+\cosh (4 r
   x)-3)}{H (2 a+\cosh (2 r x)+1)^2},\\
   p(x) &= \frac{q}{12}\left(\sigma'(x)+6p_b(x)\right)=\frac{a (4 H-3) q^2 r^2 \text{sech}^2(r x) (-4 a-2 \cosh (2 r x)+\cosh (4 r
   x)-3)}{6 H (2 a+\cosh (2 r x)+1)^2}.
\end{align}
\end{subequations}

\section{One-dimensional solitary wave solution of SGN}
For the sake of completeness, we recall the explicit formula for the one-dimensional
solitary wave solutions of SGN \cite{bristeau_energy-consistent_2015}:
  \begin{subequations}
    \label{eq: 1d soliton SGN}
\begin{align}
& \ell = \sqrt{\left(\frac{H^3}{a} + H^2\right)\frac{4}{3}},
\quad
c_0 = \nu \sqrt{\frac{3}{4}}  \frac{\ell}{H}
\sqrt{\frac{g\, H^3}{\tfrac{3}{4} \ell^2 - H^2}},
\quad 
\xi = x - c_0 t - x_0,\\ 
& {h}(x,t) = H + a \, \operatorname{sech}^2\!\left(\frac{\xi}{\ell}\right),\quad
{u}(x,t) = c_0 \left(1 - \frac{H}{{h}(x,t)}\right),\\
& {w}(x,t) = -\frac{a c_0 H}{\ell\, {h}(x,t)} \,
\operatorname{sech}\!\left(\frac{\xi}{\ell}\right)\,
\operatorname{sech}'\!\left(\frac{\xi}{\ell}\right),\\
& {p}(x,t) =
\frac{a c_0^2 H^2}{2 \ell^2 {h}(x,t)^2}
\left[
\left( (2H - {h}(x,t)) \operatorname{sech}'\!\left(\frac{\xi}{\ell}\right) \right)^2
+ {h}(x,t)\, \operatorname{sech}\!\left(\frac{\xi}{\ell}\right)\, \operatorname{sech}''\!\left(\frac{\xi}{\ell}\right)
\right],
\end{align}
  \end{subequations}
where $H$ is the depth of the ocean at rest, $a$ is the amplitude of the solitary wave, $\nu$ indicates the direction of propagation,
and $x_0$ is the initial position of the wave peak.

\section{Eigenstructure of mbHSGN and HSGN}
For the discretization of mbHSGN and HSGN used in this work, it suffices to compute the eigenstructure
of the systems with an additional passive tracer variable corresponding to the tangential velocity $v$.
For mbHSGN, the eigenvalues of 
\begin{align}
    \partial \bfF^x/\partial \bfq + \bfA^x = 
    \left(
\begin{array}{ccccc}
 0 & 1 & 0 & 0 & 0 \\
 g h-u^2 & 2 u & 0 & 0 & 1 \\
 -u v & v & u & 0 & 0 \\
 -u w & w & 0 & u & 0 \\
 c^2 (-u)-p u & c^2+p & 0 & 0 & u \\
\end{array}
\right),
\end{align}
are given by
\begin{align}
    \lambda_1 &= u - \sqrt{g h + p + c^2},\quad
    \lambda_{2,3,4} = u,\quad
    \lambda_5 = u + \sqrt{g h + p + c^2}.
\end{align}
The corresponding matrix of right eigenvectors and its inverse are given by
\begin{align}
    \bfR = \left(
\begin{array}{ccccc}
 1 & 1 & 0 & 0 & 1 \\
 u-C_e & u & 0 & 0 & C_e+u \\
 v & 0 & 0 & 1 & v \\
 w & 0 & 1 & 0 & w \\
 c^2+p & -g h & 0 & 0 & c^2+p \\
\end{array}
\right), \quad 
\bfR^{-1} = \left(
\begin{array}{ccccc}
 \frac{u C_e+g h}{2 C_e^2} & -\frac{1}{2 C_e} & 0 & 0 & \frac{1}{2
   C_e^2} \\
 \frac{c^2+p}{C_e^2} & 0 & 0 & 0 & -\frac{1}{C_e^2} \\
 -\frac{g h w}{C_e^2} & 0 & 0 & 1 & -\frac{w}{C_e^2} \\
 -\frac{g h v}{C_e^2} & 0 & 1 & 0 & -\frac{v}{C_e^2} \\
 \frac{g h-u C_e}{2 C_e^2} & \frac{1}{2 C_e} & 0 & 0 & \frac{1}{2 C_e^2}
   \\
\end{array}
\right),
\end{align}
where $C_e = \sqrt{g h + p + c^2}$ is the effective wave speed.
For HSGN, we have
\begin{align}
    \partial \bfF^x/\partial \bfq + \bfA^x = 
    \left(
\begin{array}{ccccccc}
 0 & 1 & 0 & 0 & 0 & 0 & 0 \\
 g h-u^2 & 2 u & 0 & 0 & 0 & 1 & 0 \\
 -u v & v & u & 0 & 0 & 0 & 0 \\
 -u w & w & 0 & u & 0 & 0 & 0 \\
 \sigma  (-u) & \sigma  & 0 & 0 & u & 0 & 0 \\
 c^2 (-u)-p u & c^2+p & 0 & 0 & 0 & u & 0 \\
 -u p_b & p_b & 0 & 0 & 0 & 0 & u \\
\end{array}
\right)
\end{align}
with eigenvalues
\begin{align}
    \lambda_1 &= u - \sqrt{g h + p + c^2},\quad
    \lambda_{2,3,4,5,6} = u,\quad
    \lambda_7 = u + \sqrt{g h + p + c^2}.
\end{align}
The matrix of right eigenvectors is given by
\begin{align}
    \bfR =
\left(
\begin{array}{ccccccc}
 1 & 1 & 0 & 0 & 0 & 0 & 1 \\
  u-C_e & u & 0 & 0 & 0 & 0 &
   C_e+u \\
 v & 0 & 0 & 0 & 1 & 0 & v \\
 w & 0 & 0 & 1 & 0 & 0 & w \\
 \sigma  & 0 & 1 & 0 & 0 & 0 & \sigma  \\
 c^2+p & -g h & 0 & 0 & 0 & 0 & c^2+p \\
 p_b & 0 & 0 & 0 & 0 & 1 & p_b \\
\end{array}
\right),
\end{align}
and its inverse is given by
\begin{align}
    \bfR^{-1} = 
\left(
\begin{array}{ccccccc}
 \frac{u C_e+g h}{2 C_e^2} & -\frac{1}{2 C_e} & 0 & 0 & 0 &
   \frac{1}{2 C_e^2} & 0 \\
 \frac{c^2+p}{C_e^2} & 0 & 0 & 0 & 0 & -\frac{1}{C_e^2} & 0 \\
 -\frac{g h \sigma }{C_e^2} & 0 & 0 & 0 & 1 & -\frac{\sigma }{C_e^2} & 0 \\
 -\frac{g h w}{C_e^2} & 0 & 0 & 1 & 0 & -\frac{w}{C_e^2} & 0 \\
 -\frac{g h v}{C_e^2} & 0 & 1 & 0 & 0 & -\frac{v}{C_e^2} & 0 \\
 -\frac{g h p_b}{C_e^2} & 0 & 0 & 0 & 0 & -\frac{p_b}{C_e^2} & 1 \\
 \frac{g h-u C_e}{2 C_e^2} & \frac{1}{2 C_e} & 0 & 0 & 0 &
   \frac{1}{2 C_e^2} & 0 \\
\end{array}
\right).
\end{align}

\section{Radially symmetric two-dimensional solution of mbHSGN and HSGN}
If we consider radially symmetric solutions of mbHSGN and HSGN, the 2D systems collapse to the 1D systems,
replacing the Cartesian coordinate $x$ by the radial coordinate $r$, the velocity component $u$ representing the radial velocity,
and with the addition of forcing source terms.
Such source terms take the form:
\begin{gather}
    \Tilde{\bfS}_{\text{mbHSGN}}(\bfq,r):=-\frac{1}{r}
    \left(
    h u, h u^2, h u w, h u (p+c^2)
    \right)^T,\\
    \Tilde{\bfS}_{\text{HSGN}}(\bfq,r):=-\frac{1}{r}
    \left(
    h u, h u^2, h u w, h u \sigma, h u (p+c^2), h u p_b
    \right)^T,
\end{gather}
for the mbHSGN and
HSGN systems respectively.

\section{Transition to SWE versus bathymetry effects}
We provide some numerical evidence supporting the claim that a sharp transition to the SWE
seems to have a significantly smaller impact on the surface elevation than bathymetry effects.
Of course, this is a very preliminary observation, and a more systematic and quantitative study
would be needed to draw more definitive conclusions.

\begin{figure}
    \centering
    \includegraphics[width=\textwidth]{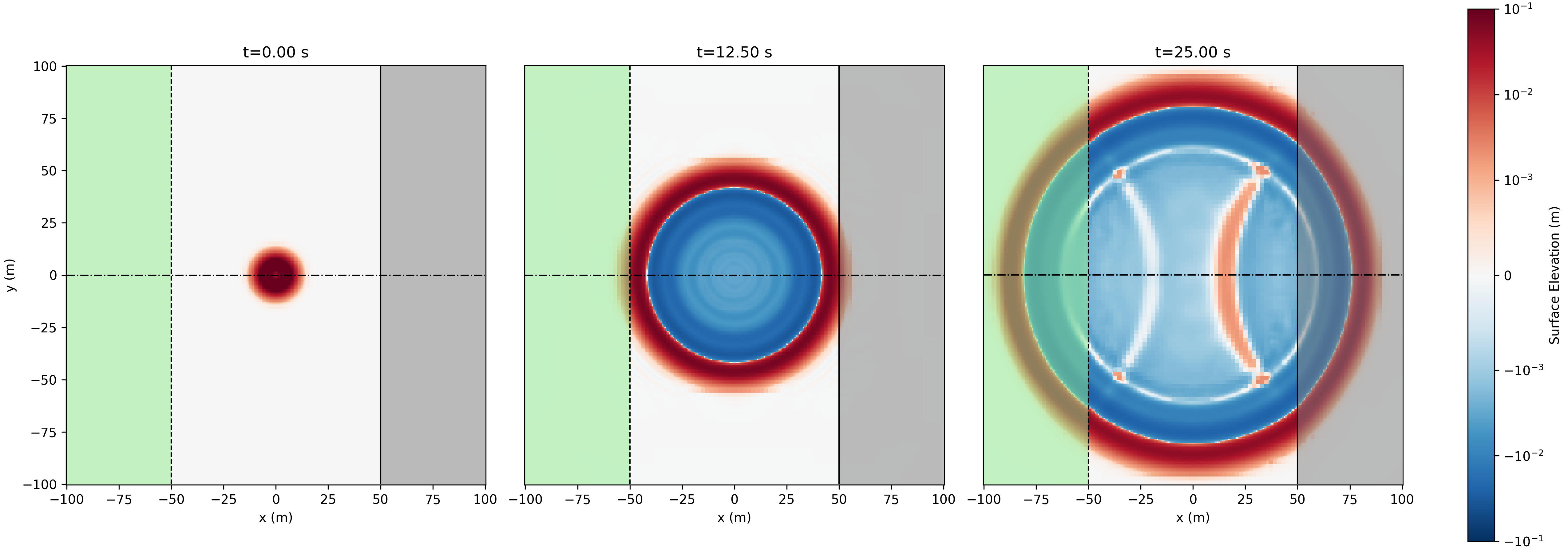}
    \caption{Surface elevation at three different times for the mbHSGN model. The light green region corresponds to the region where the transition to SWE occurs,
    the light gray region corresponds to the region where the bathymetry jump occurs ($B/H = 0.3$). \label{fig: 2d plot transition bathymetry}}
\end{figure}

\begin{figure}
    \centering
    \includegraphics[width=\textwidth]{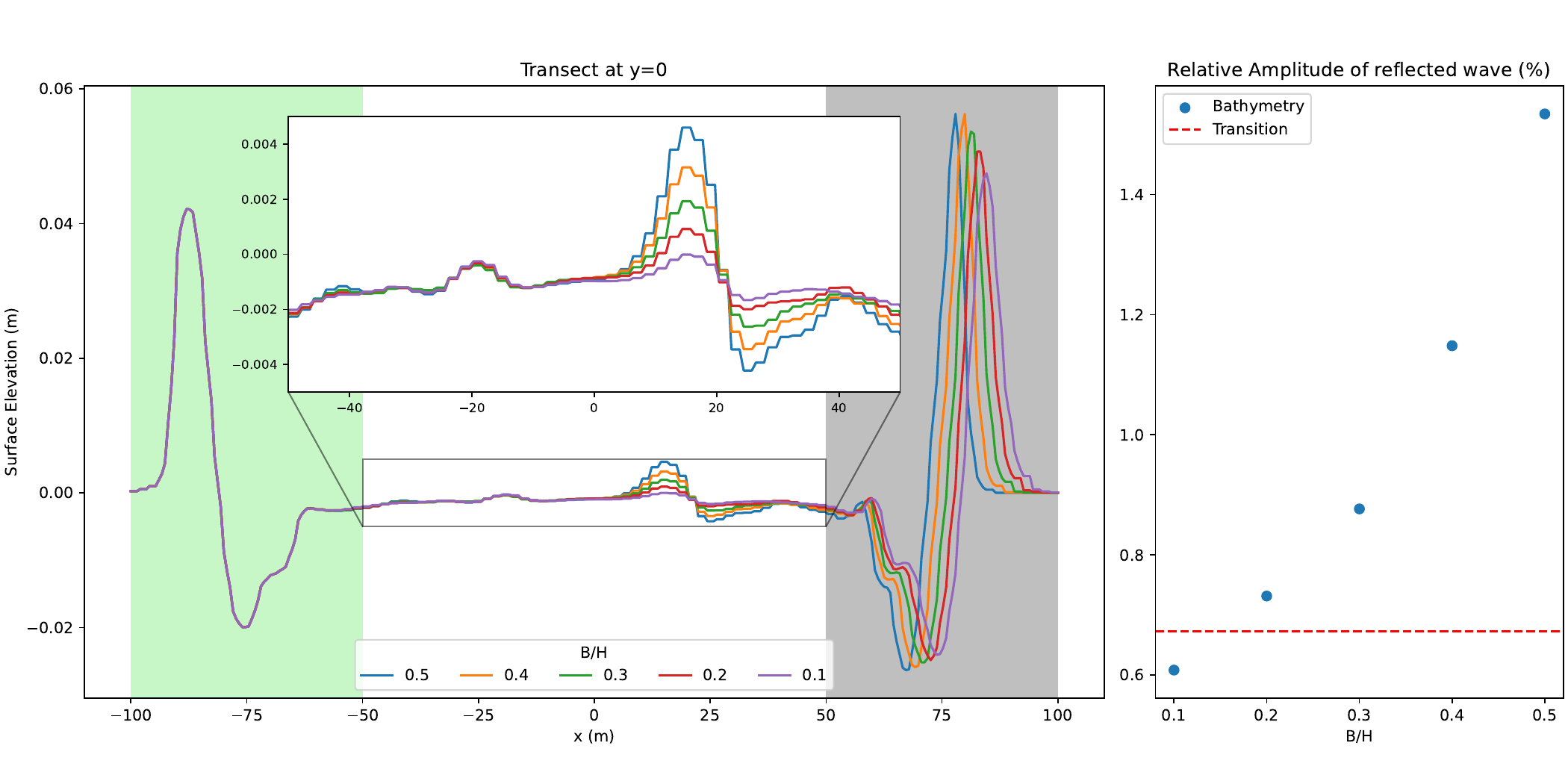}
    \caption{Left: Surface elevation along the transect $y=0$ at $t=100 \si{\second}$ for the mbHSGN model. The light green region corresponds to the region where the transition to SWE occurs,
    the light gray region corresponds to the region where the bathymetry jump occurs (different values of $B/H$).
    Right: Scatter plot of maximum amplitude in the bathymetry region vs. maximum amplitude in the transition region. \label{fig: 1d plot transition bathymetry}}
\end{figure}

We consider a solitary wave of SGN (with $a=0.3 \si{\meter}$ and $H=1 \si{\meter}$) in the radial direction
and discretize mbHSGN (with $\alpha=3$)
on a grid of size $[-100,100]\times[-100,100]$. 
We force a transition to the SWE at the left of $x=-50 \si{\meter}$, and we introduce a 
bathymetry jump of different sizes at $x=50 \si{\meter}$.
The surface elevation at three different times is shown in Figure
\ref{fig: 2d plot transition bathymetry}, where the light green region corresponds to the region where the transition to SWE occurs, 
and the light gray region corresponds to the region where the bathymetry jump occurs.
The size of the bathymetry jump is given by different values of $B/H\in \{0.1, 0.2, 0.3, 0.4, 0.5\}$,
where $B$ is the jump in bathymetry and $H$ is the reference water depth.
In Figure \ref{fig: 1d plot transition bathymetry} we can see that the reflections caused by the bathymetry are larger than those caused by the transition to the SWE
up to the point where the bathymetry jump is around $10 \si{\percent}$ of the reference water depth $H$ (noting that
the amplitude of the reflected wave from the transition is less than $1 \si{\percent}$ of the amplitude of the incoming wave).


\section{Effect of the bathymetry sharpness sensor}

\begin{figure}
    \centering
    \includegraphics[width=\textwidth]{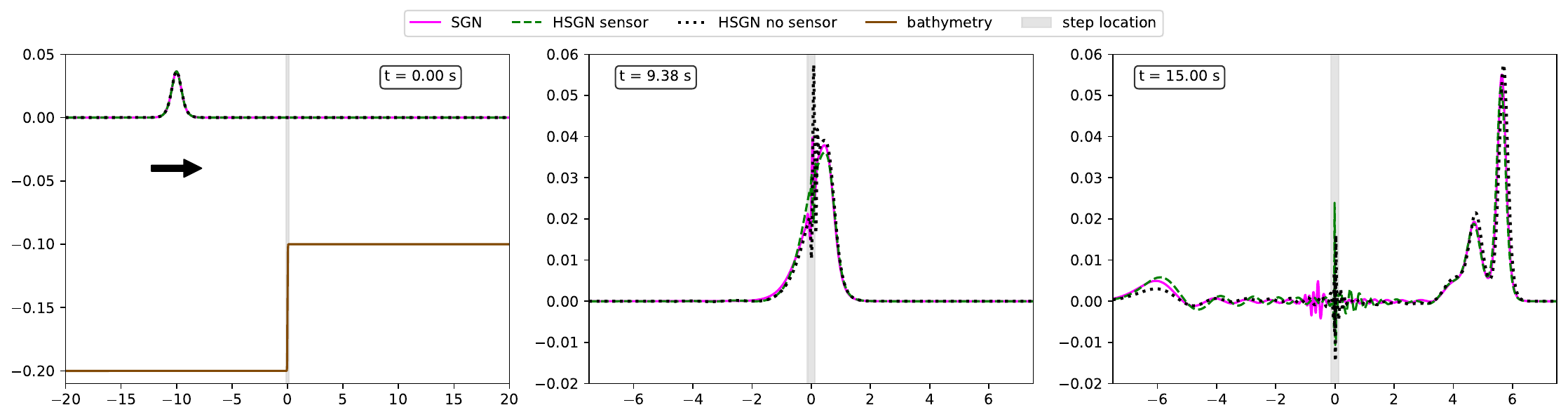}
    \caption{Surface elevation at three different times for HSGN  with and without the bathymetry sharpness sensor. The light gray region corresponds to the region where the bathymetry jump occurs. 
    $N=8000$ cells are used ($\Delta=0.0075 \si{\meter}$)\label{fig: sensor effect different times}}
\end{figure}

\begin{figure}
    \centering
    \includegraphics[width=\textwidth]{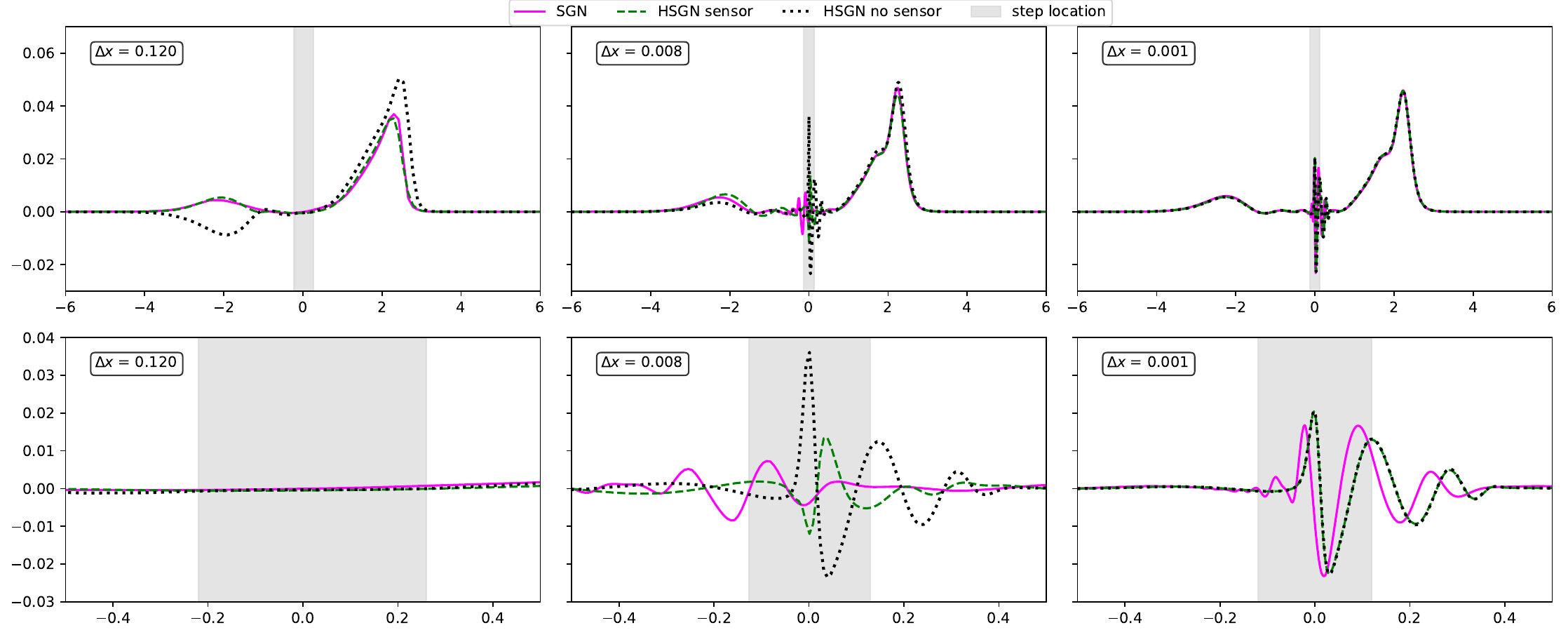}
    \caption{Surface elevation at $t =25 \si{\second}$ for increasingly refined grids. The light gray region corresponds to the region where the bathymetry jump occurs. \label{fig: sensor effect different mx}}
\end{figure}

To illustrate the type of behavior that the bathymetry sharpness sensor is designed to capture, we consider a setup analogous
to the one used in \cite[Sec. 4.2]{bassi_hyperbolic_2020},
where 
a one-dimensional
solitary wave solution of SGN propagates over a smoothed bathymetry step given by the
error function  
\begin{align}
b(x) = 0.05 ( \mathrm{erf}(32x) + 1 )-H,
\end{align}
with $H=0.2 \si{\meter}$ being the reference water depth.
The setup and initial conditions are  depicted in the left panel of Figure \ref{fig: sensor effect different times},
while the surface elevation at different times is shown in the center and right panels of the same figure, where the light gray region corresponds to the region where the bathymetry jump occurs.
As the soliton propagates over the bathymetry step, some high-frequency oscillations are generated
over the bathymetry jump. 
If under-resolved, these oscillations can lead to numerical instabilities.
Thus, the bathymetry sharpness sensor is designed to suppress such oscillations
until the bathymetric effects are sufficiently resolved,  and recover a consistent discretization on smooth bathymetries and fine grids.
This is illustrated in Figure \ref{fig: sensor effect different mx}.

While these oscillations seem to be exacerbated by the hyperbolic relaxation, 
we have consistently observed them in simulations with SGN in extremely resolved grids.
A careful investigation of the nature and characterization of these oscillations is part of ongoing work.

\end{document}